\documentclass[sigconf, nonacm]{acmart}
\usepackage{mathtools}
\usepackage{textcomp}
\usepackage{fancyhdr,graphicx,amsmath}
\usepackage[ruled,vlined,linesnumbered]{algorithm2e}
\usepackage{amsmath}
\usepackage[capitalize,noabbrev]{cleveref}

\usepackage[textsize=tiny]{todonotes}
\usepackage{hyperref}
\usepackage{enumitem}
\usepackage{tabu}
\usepackage{amsmath,amsfonts}
\usepackage{xspace}
\usepackage{textcomp}
\usepackage[utf8]{inputenc}
\usepackage[T1]{fontenc}
\usepackage{microtype}
\usepackage[a-2b]{pdfx}
\usepackage{graphicx}
\usepackage{balance}
\usepackage{subcaption}
\usepackage{multirow}
\usepackage{xurl}
\usepackage{amsthm}

\SetCommentSty{mycommfont}
\usepackage{color}
\usepackage{listings}
\usepackage{float}
\usepackage{xcolor}
\usepackage{outlines}
\usepackage[normalem]{ulem}
\usepackage{cleveref}
\usepackage{soul}
\usepackage{cuted, lipsum}
\usepackage[listings,skins,breakable]{tcolorbox}
\usepackage{booktabs}
\usepackage{float}
\usepackage{placeins}
\usepackage{hyperref}

\newcommand{\eat}[1]{}

\newcommand{\ExBoost}{\textsc{InferF}\xspace}

\newcommand{\InferF}{\textsc{InferF}\xspace}

\lstdefinestyle{sqlstyle}{
    language=SQL,
    basicstyle=\ttfamily\small,
    keywordstyle=\color{blue},
    commentstyle=\color{gray},
    stringstyle=\color{red},
    numbers=left,
    numberstyle=\tiny\color{gray},
    stepnumber=1,
    numbersep=5pt,
    showstringspaces=false,
    breaklines=true,
    frame=single,
    captionpos=b
}

\lstdefinestyle{pythonstyle}{
    language=Python,
    basicstyle=\ttfamily\small,
    keywordstyle=\color{blue},
    commentstyle=\color{gray},
    stringstyle=\color{red},
    numbers=left,
    numberstyle=\tiny\color{gray},
    stepnumber=1,
    numbersep=5pt,
    showstringspaces=false,
    breaklines=true,
    frame=single,
    captionpos=b
}

\lstdefinestyle{felstyle}{
    language=Python,
    basicstyle=\ttfamily\small,
    keywordstyle=\color{blue},
    commentstyle=\color{gray},
    stringstyle=\color{red},
    numbers=left,
    numberstyle=\tiny\color{gray},
    stepnumber=1,
    numbersep=5pt,
    showstringspaces=false,
    breaklines=true,
    frame=single,
    captionpos=b
}

\lstnewenvironment{DSL}
  {\lstset{
        aboveskip=5pt,
        belowskip=5pt,
        mathescape=true,
        basicstyle=\ttfamily\small,
        morekeywords={Set, Vector, Map, HashMap, bool, project,
  filter, aggregate, join, partition, member, method, construct, true, false, if, else, CREATE TABLE, PARTITION BY,
  self, literal, vector, return, for push_back, function, enum, sort},
        literate={~} {$\sim$}{1},
        showstringspaces=false}\vspace{0pt}%
   \noindent\minipage{0.47\textwidth}}
   {\endminipage\vspace{0pt}}

\lstnewenvironment{SQL}
  {\lstset{
        aboveskip=5pt,
        belowskip=5pt,
        escapechar=!,
        mathescape=true,
        basicstyle=\linespread{0.94}\ttfamily\small,
        morekeywords={FOR, EACH, WITH, CREATE, TABLE, BY, HASH, PARTITION, TEST, WHETHER, PROBABILITY},
        deletekeywords={VALUE, PRIOR},
        showstringspaces=false}
        \vspace{0pt}%
        \noindent\minipage{0.47\textwidth}}
  {\endminipage\vspace{0pt}}

\newcommand{\lixi}[1]{\textcolor{purple}{(Lixi:#1)}}

\newcommand{\revision}[1]{\textcolor{blue}{#1}}

\newcommand{\littlesection}[1]

\AtBeginDocument{%
  }

\setcopyright{acmlicensed}
\copyrightyear{2018}
\acmYear{2018}
\acmDOI{XXXXXXX.XXXXXXX}
\acmConference[Conference acronym 'XX]{Make sure to enter the correct
  conference title from your rights confirmation email}{June 03--05,
  2018}{Woodstock, NY}
\acmISBN{978-1-4503-XXXX-X/2018/06}

\begin{document}

\title[\InferF: Declarative Factorization of AI/ML Inferences over Joins]{\InferF: Declarative Factorization of AI/ML Inferences over Joins \\\Large Accepted to SIGMOD 2026 as a Full Research Paper}

\author{Kanchan Chowdhury$^{a,c}$, Lixi Zhou$^a$, Lulu Xie$^a$, Xinwei Fu$^b$, Jia Zou$^a$ }

 \affiliation{
   \institution{$^a$ Arizona State University, $^b$ Amazon, $^c$ Marquette University}
  \country{}
   }

\email{kanchan.chowdhury@marquette.edu, {lixi.zhou, luluxie}@asu.edu, fuxinwe@amazon.com, jia.zou@asu.edu}

\thanks{Most of Kanchan Chowdhury's work is done when he was in ASU; Jia Zou is the corresponding author.}

\begin{abstract}
Real-world AI/ML workflows often apply inference computations to feature vectors joined from multiple datasets. To avoid the redundant AI/ML computations caused by repeated data records in the join's output, \textit{factorized} ML has been proposed to decompose ML computations into sub-computations to be executed on each normalized dataset. \textcolor{black}{However, there is insufficient discussion on how \textit{factorized} ML could impact AI/ML inference over multi-way joins. To address the limitations, we propose a novel declarative \InferF system, focusing on the factorization of arbitrary inference workflows represented as analyzable expressions over the multi-way joins. We formalize our problem to flexibly push down partial factorized computations to qualified nodes in the join tree to minimize the overall inference computation and join costs and propose two algorithms to resolve the problem:} (1) a greedy algorithm based on a per-node cost function that estimates the influence on overall latency if a subset of factorized computations is pushed to a node, and (2) a genetic algorithm for iteratively enumerating and evaluating promising factorization plans. \textcolor{black}{We implement \InferF on Velox, an open-sourced database engine from Meta, evaluate it on real-world datasets, observed up to \textbf{$11.3\times$} speedups, and systematically summarized the factors that determine when \textit{factorized} ML can benefit AI/ML inference workflows.}
\end{abstract}

\maketitle

\section{Introduction}
\label{sec:introduction}
Running AI/ML inferences over data denormalized from multiple tables is a prevalent pattern in data science pipelines. Out of $2{,}000$ most recent Jupyter notebooks involving at least one join, which we crawled from Github, $18.6\%$ of them contain $6$ to $101$ joins. Moreover, among the $173{,}851$ SQL files in \textit{The Stack}~\cite{Kocetkov2022TheStack} that contain at least one join and represent feature processing and other analytics tasks, $29.6\%$ involve $6$ to $3{,}573$ joins.

 
Data denormalization introduces data redundancy, causing redundant AI/ML computations. For example, the normalized IMDB tables require only $1.2$ gigabytes of storage, whereas the fully denormalized IMDB tables exceed $1$ terabyte~\cite{factorize-joinboost}\textcolor{black}{, representing nearly a $1000{\times}$ increase due to redundant data replication across joins, which leads to redundant (sub-)inference computations}.  To address the problem, factorized ML~\cite{factorize-join, kumar2015learning, factorize-lmfao, factorize-joinboost, factorize-la, li2019enabling} is proposed to run the end-to-end processing in the database, decompose the ML computations, and push the decomposed computations down through the join operators. However, most of the existing factorized ML works focus on the AI/ML training process, including F~\cite{factorize-join} for regression models, FL~\cite{kumar2015learning} for generalized linear models, LMFAO~\cite{factorize-lmfao} for decision trees, Bayesian networks, etc., JoinBoost~\cite{factorize-joinboost} for gradient boosting trees, Morpheus~\cite{factorize-la} and MorpheusFI~\cite{li2019enabling} for linear algebra operators in training workloads.
\textcolor{black}{None of the existing in-database factorized ML systems~\cite{factorize-lmfao, factorize-f, factorize-join, factorize-joinboost, factorize-ac-dc, factorize-faq-ai, factorize-database, kumar2015learning, factorize-la, li2019enabling, shaikhha2022functional} have systematically explored the potential of factorized ML for the inference process. In particular, considering the diversity of data and join cardinalities in multi-way join queries, fine-grained factorization of inference computations and selective (group) push-down strategies could yield substantial performance improvements, which haven't been discussed in any existing works. Furthermore, existing systems primarily focus on factorizing specific ML models, operators, or transformed semirings, making them difficult to generalize to arbitrary inference workflows. 
}

\vspace{5pt}
To bridge the gap, we first formalize the  problem of optimizing AI/ML inference computations over multi-way joins without affecting inference accuracy. \textcolor{black}{In contrast to existing works, which focus on the training processes, our system focuses on factorizing the model inference process. We aim to accelerate the inference of pretrained models over relational data.}
In our target scenarios, the testing dataset to be scored is first prepared using feature extraction queries, which are then applied with AI/ML inference functions for popular models such as deep neural networks, decision trees, and nearest neighbor search, to drive insights and decision-making. This scenario is ubiquitous in data science pipelines and enterprise ML applications\textcolor{black}{~\cite{armenatzoglou2022amazon, bisong2019google, boehm2016systemml, boehm2019systemds, gaussml, guan2023comparison, DBLP:conf/edbt/ZhouLCMEMSWWWY024, DBLP:journals/pvldb/ZhouCDMYZZ22, guan2025privacy, zhou2024deepmapping}}. For example, in $130$ customer engagements with Microsoft, the requirements of
$91\%$ of them were captured by such batch inference~\cite{park2022end}. Additionally, our work can be applied to more complex inference queries where AI/ML-based filter predicates and projections are nested within SQL~\cite{park2022end, zhang2025mitigating}. 
It is critical to reduce the end-to-end latency of inference queries to improve user experiences and reduce operational costs, since inference queries drive up to $90\%$ of costs associated with ML in the enterprise~\cite{amazon-keynote, amazon-tco, crankshaw2017clipper, shen2019nexus}. 

Our work focuses on addressing the following {\textit{challenges}}:

\noindent
\textbf{Challenge-1: Exponential Search Space.}
Flexibly assigning each group of factorized arbitrary (sub-)computations to datasets or  \textcolor{black}{join} nodes to minimize execution costs is a new problem, which faces an exponential search space, as analyzed in Sec.~\ref{sec:np-hard}.

\noindent
\textbf{Challenge-2: Cost Estimation.} Given the diverse inference workflows with different
complexity and output dimension size of each factorized sub-computation, different number and cardinalities of join operators, and different sizes of underlying datasets, it is hard to understand the key trade-offs that determine the benefits and costs of each fine-grained factorization and push-down plan.

\noindent
\textbf{Challenge-3: Arbitrary Inference Workflows.} AI/ML inference computations are often encapsulated in UDFs that are opaque to query optimizers, making it hard to analyze which portions of inference workflows are factorizable, and which tables or intermediate joins the computations could be pushed to. 


\vspace{3pt}
\textcolor{black}{To the best of our knowledge, no existing in-DB factorized ML system has addressed the aforementioned challenges.}
In this work, to address the challenges, we propose a declarative system called \InferF with the following \textbf{key contributions}:

\vspace{3pt}
\noindent
\textit{(1) A New Problem Formalization.}  {\textcolor{black} {Prior systems focus on abstracting the training process as a batch of aggregation queries (e.g., LMFAO~\cite{factorize-lmfao, factorize-ac-dc, factorize-f, factorize-join}), a semi-ring~\cite{factorize-joinboost, shaikhha2022functional}, or iterative linear algebra workflows~\cite{kumar2015learning, factorize-la, li2019enabling} and introduce rules to determine ``how to factorize'' for training using specific learning algorithms. None of these works formalize optimization problems on how to factorize the underlying abstraction to achieve the best performance. Unlike them, we provide the \textbf{first formal definition of the fine-grained factorization optimization problem for general inference workflows} over multi-way joins. Our formulation supports \textbf{group push-down} for aggressive aggregation of intermediate inference results, proves NP-hardness, and shows that factorization optimization can be safely \textbf{decoupled from join-order optimization}, enabling principled reasoning beyond model-specific or rule-based strategies. Our formulation has no assumptions on model types, join order,  join algorithms, or table/feature ratios. }}

\vspace{3pt}
\noindent
\textit{(2) New Optimization Algorithms.}
\textcolor{black}{Existing factorized ML systems rely on simplified decision rules based on counts of GroupBy attributes in each view (e.g., LMFAO~\cite{factorize-lmfao}), tuple/feature ratio (e.g., Morpheus~\cite{factorize-la}), or cost analyses specific to iterative batch gradient descent (e.g., FL~\cite{kumar2015learning}), making training-specific decisions such as selecting a view for each aggregation query~\cite{factorize-lmfao, factorize-joinboost}, or coarse-grained decisions such as determining whether to perform factorization~\cite{kumar2015learning, factorize-la}. None of them can be applied to factorize inference computations that are not iterative and do not involve feature-level aggregations to make fine-grained decisions that assign each factorized sub-computation to a node in a join tree. 
In contrast, we introduce two \textbf{cost-driven, model-agnostic optimization algorithms} that systematically explore the exponential search space of the fine-grained factorization and push-down plans for inference workflows. 
The first is a \textbf{genetic evolutionary optimizer}~\cite{bennett1991genetic} that operates at the plan level, using our unified cost model to evolve candidate push-down plans through mutation and crossover while balancing computation and I/O trade-offs. 
The second is a \textbf{greedy benefit-first optimizer} that iteratively selects the next best node in a join tree for factorized computation based on a lightweight node-level cost function combining \textbf{Sobol sensitivity analysis}~\cite{sobol-1,sobol-2} with a learned \textbf{logistic regression benefit predictor}. 
Together, these algorithms move beyond heuristic or rule-based optimization, offering the first principled framework for cost-based inference factorization.}

\vspace{3pt}
\noindent
\textit{(3) A New Factorization Framework in UDF-Centric Database.} We allow relational algebra operators to be customized by AI/ML logics that are described by an analyzable expression. 
The expression represents a directed acyclic computational graph, where a node can be an expression operator, such as arithmetic operations, relation operations, bit operations, conditional operations, and atomic user-defined functions (UDFs). The expression graph will be analyzed to identify subgraphs that only rely on one source relation independently and could be factorized and pushed down. 

\vspace{3pt}
\noindent
\textit{(4) Implementation, Evaluation, and Insights.} Finally, we implemented the \InferF system on Velox~\cite{pedreira2022velox}, a high-performance, UDF-centric database engine open-sourced from Meta. We conducted an experimental evaluation on synthetic and real-world inference query workloads to understand the effectiveness of the proposed \InferF approach across diverse use scenarios. Our evaluation results show that our proposed factorized ML approach achieved up to \textcolor{black}{$\textbf{11.3}\times$} speedup over the best baseline within Velox, and outperforms the best of other in-DB ML systems by up to $\textbf{18.7}\times$. We also provide a list of empirically verified observations about whether and when factorized ML will benefit end-to-end inference workflows.

To our knowledge, \InferF is the first system to address end-to-end factorization optimization of AI/ML inference over multi-way joins, especially for opaque and non-linear inference workflows. It bridges a practical gap between modern inference workloads and database optimization strategies, opening new possibilities for cost-effective, production-scale AI/ML inference.

\vspace{-5pt}
\section{Background}
\label{sec:back-related-work}
\textcolor{black}{In this section, we first demonstrate the lossless factorization of the inference processes for pre-trained models—such as FFNNs, decision trees, and product quantization—as well as user-defined inference workflows over features derived from multi-way relational joins. We then present its potential benefits.}

\vspace{-5pt}
\subsection{Factorization of Inference Processes}
\label{sec:examples}
\subsubsection{FFNN}
\label{sec:dnn}

Taking a fully connected (FC) layer with the weight matrix $\boldsymbol{W}$ of shape $n \times k$ as an example, the layer will convert an $n$-dimensional input feature vector $\boldsymbol{x}$ into $k$ hidden features (or outputs) using $\sigma(\boldsymbol{W^T} \boldsymbol{x} + \boldsymbol{b})$, where $\sigma$ represents the activation function such as ReLU and $\boldsymbol{b}$ represents a bias vector. We suppose the features sent to the layer are created by an $m$-way join over $R_1$,...,$R_m$, so that a sub-feature-vector $\boldsymbol{x_i}$ is derived from $R_i$. Then, $\boldsymbol{W}$ can be co-partitioned into $m$ submatrices corresponding to sub-feature-vectors $\boldsymbol{x_1}$,..., $\boldsymbol{x_m}$, so that the $i$-th partition $\boldsymbol{W_i}$ has a shape of $ d(\boldsymbol{x_i}) \times k$, where $d(\boldsymbol{x_i})$ represents the number of features in $\boldsymbol{x_i}$. $\boldsymbol{W_i}$ represents the weight parameters at all edges that connect the features of $\boldsymbol{x_i}$ and all $k$ hidden/output neurons. Then, we have factorization $  \boldsymbol{W^T} \boldsymbol{x} = \Sigma_{i=1}^{m} \boldsymbol{W_i^T} \boldsymbol{x_i}$. 
Each component $\boldsymbol{W_i^T}\boldsymbol{x_i}$ can be pushed down to be computed over the \texttt{scan} node corresponding to $R_i$ or any of its ancestor join nodes. In addition, the computation of $\Sigma_{i=u}^{v}  
\boldsymbol{W_i^T}\times\boldsymbol{x_i}$ can be pushed down to the output of $R_u\bowtie ... \bowtie R_v$, which is the joined output of $R_u$, ..., $R_v$, where $1\leq u\leq v\leq m$. 

\begin{figure}[h]
\centering{%
   \includegraphics[width=2.8in]{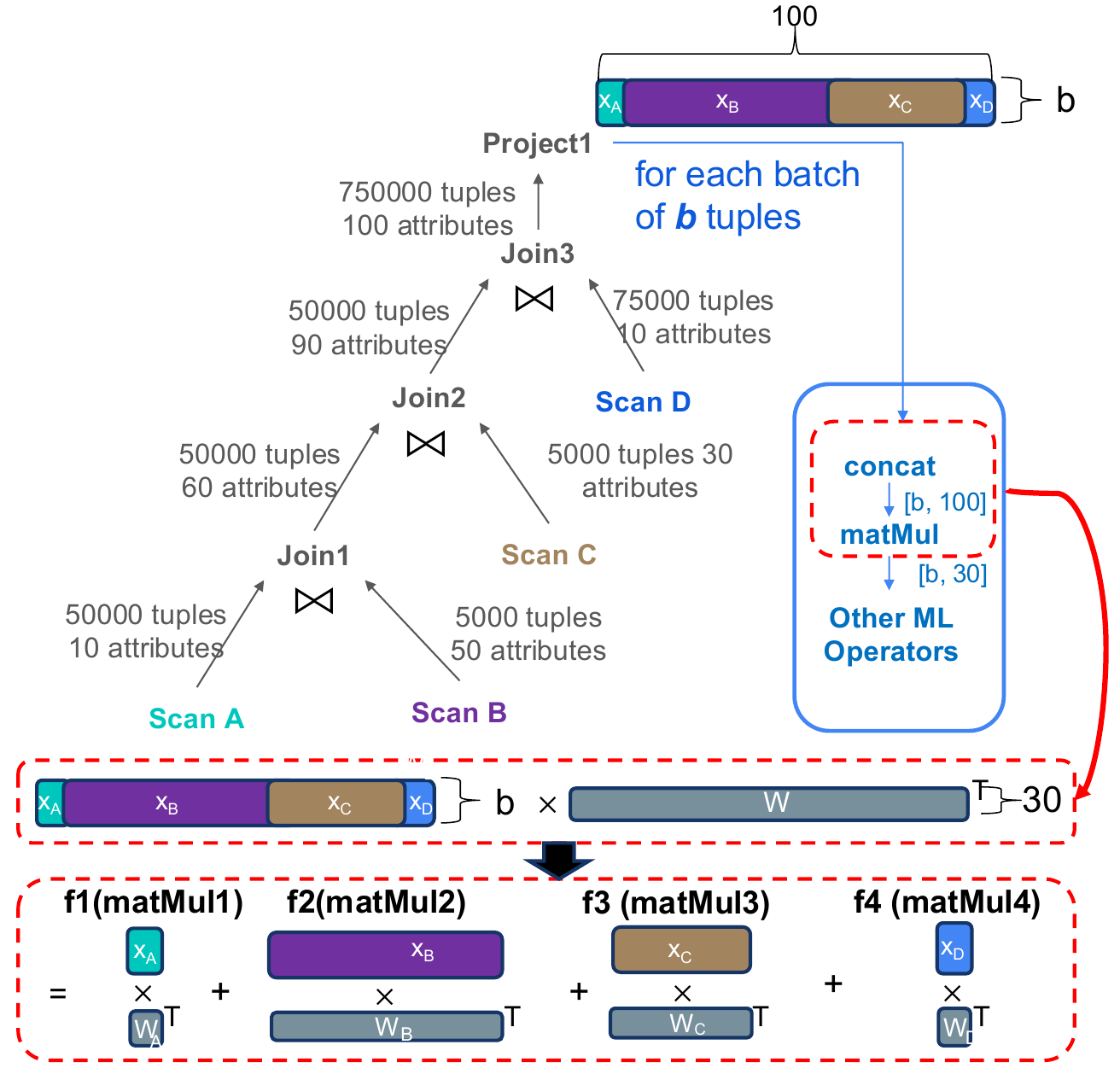}  
}
\vspace{-15pt}
\caption{\label{fig:example-1} \small {Illustration of Factorized FFNN Inference}}

\end{figure}

Fig.~\ref{fig:example-1} illustrates an example of factorizing the first matrix\_multiply (\texttt{matMul}) of a two-layer FFNN into four smaller \texttt{matMul} operators, \texttt{matMul1} (denoted $f_1$), \texttt{matMul2} ($f_2$), \texttt{matMul3} ($f_3$), and \texttt{matMul4} ($f_4$), with each multiplying a subvector of $\boldsymbol{x}$ associated with one of source datasets (A, B, C, and D) to the corresponding partition of weight matrix $\boldsymbol{W}$. $f_1$, $f_2$, $f_3$, and $f_4$ can be pushed to Scan A, Scan B, Scan C, and Scan D, as well as their ancestor nodes, respectively.

\subsubsection{Decision Tree}
\label{sec:decision-tree}

 Then, we consider a decision tree $T$ implemented using the QuickScorer algorithm~\cite{lucchese2015quickscorer, lucchese2017quickscorer}. First, a decision tree $T$ consists of (1) leaf nodes $TL=\{l_i\}$, and (2) intermediate nodes $TI=\{n_i=(fid, t_i, v_i)\}$, where $fid, t_i$ represents the feature index and the threshold of the $i$-th tree node $n_i$ respectively, forming a predicate $p_i: \boldsymbol{x}[fid] < t_i$. 
 For the tree node $n_i$, $v_i$ is a bit vector encoded for $n_i$, where each bit corresponds to a leaf node. If $v_i[d]=1$, it means that if $n_i$ is a false node (i.e., $p_i(\boldsymbol{x})=False$), the $d$-th leaf node $l_d$ is possible to be the exit leaf, and if $v_i[d]=0$, $l_d$ is impossible to be the exit leaf if $n_i$ is a false node. 
 The index of the exit leaf of the decision tree will be the index of the leftmost non-zero digit (i.e., number of trailing zeros) of $\underset{ p_i(\boldsymbol{x})=False \wedge n_i\in TI}{\wedge} v_i$. The details are described in the QuickScorer algorithm~\cite{lucchese2015quickscorer, lucchese2017quickscorer}.

 Based on the above algorithm, supposing the features input to the decision tree are the output of an $m$-way \texttt{join} of $R_1$,...$R_m$, i.e., $\boldsymbol{x}=concat(\boldsymbol{x_1},...,\boldsymbol{x_m})$, where $\boldsymbol{x_i}$ corresponds to $R_i$. 
 Then, if the nodes in $TI$ are partitioned into $m$ node groups $TI_1$,...,$TI_m$, so that all nodes in the $i$-th group have predicates involving only features from $R_i$, the function  
 $dt'(\boldsymbol{x}, TI, TL) = \underset{p_i(\boldsymbol{x})=False \wedge n_i \in TI}{\wedge}{v_i}$ can be factorized into $dt'(\boldsymbol{x}, TI,$ $ TL)= 
  dt'(\boldsymbol{x_1}, TI_1, TL) \wedge ... \wedge dt'(\boldsymbol{x_m}, TI_m, TL)$. 
  The final prediction $dt(\boldsymbol{x}, TI, TL)=numTrailing-$ $Zeros(dt'(\boldsymbol{x}, TI, TL))$. 
  Similar to DNN (Sec.~\ref{sec:dnn}), each sub-computation $dt'(\boldsymbol{x_i}, TI_i, TL)$ can be pushed down to the scan node of $TI_i$ or its ancestor nodes.

\subsubsection{Product Quantization}
\label{sec:pq}

Finally, we give an example of product quantization~\cite{prod-quant}, which is a popular nearest neighbor search algorithm for high-dimensional vectors. Given a query vector $\boldsymbol{x}$ joined from $m$ relations $R_1$,...$R_m$, we can efficiently compute the approximate distance between $\boldsymbol{x}$ and any database vector $\boldsymbol{v_i} \in V$, leveraging a precomputed centroid matrix $\boldsymbol{W}$ and encoding of $\boldsymbol{v_i}$. To obtain the centroid matrix $\boldsymbol{W}$, we first partition $v_i\in V$ into $m$ parts. Then, for the $k$-th part, we have $n$ subvectors and cluster them into $l$ clusters, and obtain the centroid of each cluster. Therefore, $\boldsymbol{W}[i][j]$ represents the centroid of the $j$-th cluster associated with the $i$-th partition. Each vector in the database $v_i\in V$ is encoded as $(c^i_1, ..., c^i_l)$, where $c^i_k$ represents the ID of the cluster to which the $v_i$'s $k$-th subvector belongs. 

The process of computing the approximate distance between a query vector $\boldsymbol{x}$ and a database vector $\boldsymbol{v_i}$ is abstracted in three steps: (1) Partition $\boldsymbol{x}$ into $m$ parts $\boldsymbol{x_1}$, ..., $\boldsymbol{x_m}$ corresponding to $R_1$,...,$R_m$; (2) For the $k$-th part, compute the partial approximate distance $d_{ik}$ between $\boldsymbol{x_k}$ and the $k$-th part of $v_i$ by using the centroid of the cluster $c^i_k$ to replace the actual $k$-th part of $v_i$; (3) return $approxDistance(\boldsymbol{x}, \boldsymbol{v_i}, \boldsymbol{W})=\sqrt{\sum\limits_{k=1}^m{{approxDistance(\boldsymbol{x_k}, \boldsymbol{W}[k][c^i_k])}^2}}$ =$\sqrt{\sum\limits_{k=1}^m{{approxDistance(\boldsymbol{x_k}, \boldsymbol{v_{ik}}, \boldsymbol{W}_k)}^2}}$, where $\boldsymbol{v_{ik}}$ is the $k$-th partition of $v_i$ and $\boldsymbol{W}_k$ is the $k$-th partition of $\boldsymbol{W}$. 

Here, a sub-computation $approxDistance(\boldsymbol{x_k}, \boldsymbol{v_{ik}}, \boldsymbol{W}_k)$ can be pushed down to the scan node of $R_k$ or any of its ancestors.

\subsubsection{User-Defined Inference Workflows}

\begin{figure}[t]
\centering{%
   \includegraphics[width=2.3in]{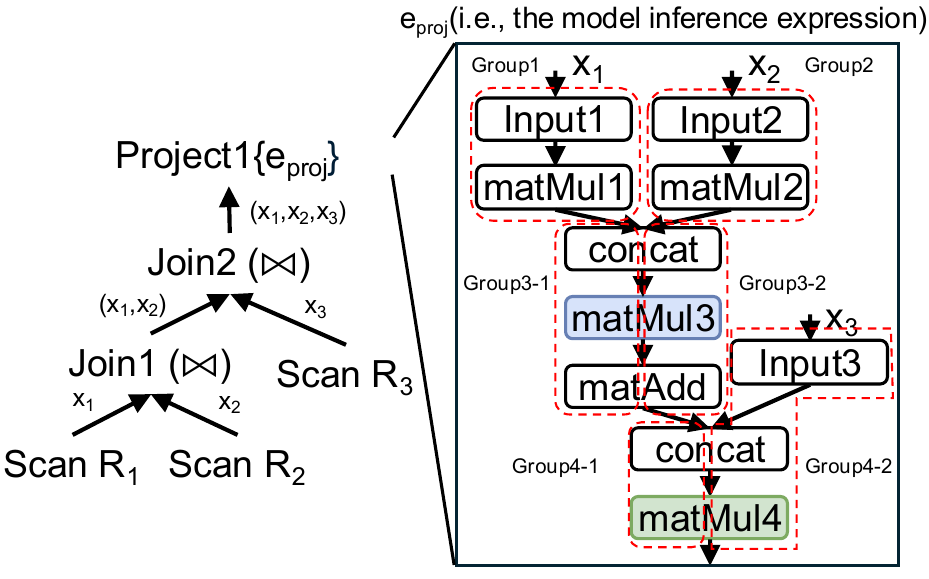}  
}
\vspace{-10pt}
\caption{\label{fig:user-defined-workflow} \small User-Defined Inference Workflow}
\end{figure}

\textcolor{black}{Fig.~\ref{fig:user-defined-workflow} illustrates a user-defined inference workflow that applies the model inference (detailed in a computational graph, as illustrated in the right part of Fig.~\ref{fig:user-defined-workflow}) over features integrated by a \textit{3}-way join. The query can be optimized with factorization and pushdown as follows: (1) Group1 and Group2 in the computational graph of the model inference can be factorized from the computational graph, and pushed down to the scan nodes over $R_1$ and $R_2$, respectively, or their ancestor join node Join1. (2) MatMul3 can be factorized into Group3-1 and Group3-2, which can also be pushed down to scan nodes of $R_1$ and $R_2$, respectively, or their ancestor join nodes Join1. (3) MatMul4 can also be factorized into Group4-1 and Group4-2.  Group4-1 can be pushed to the output of Join1, while Group4-2 can be pushed down to the scan node over $R_3$.}

{\color{black}{
\subsection{Motivations}
\label{sec:benefits}

Performing inference computation factorization and push-down yields two key benefits.
\textbf{(1) Reduced CPU cost.}
Factorizing and pushing down inference sub-computations to their corresponding source (normalized) tables eliminates redundant computations arising from duplicate feature values in join outputs. 
However, pushing every sub-computation to base tables is not always optimal: if some base tables have higher cardinalities than intermediate or root join nodes, this may increase the total computation cost. 
Therefore, an effective plan must determine the optimal granularity of factorization and the appropriate node in the join tree for each sub-computation to minimize overall CPU cost.
\textbf{(2) Reduced I/O cost.}
Pushing down a factorized sub-computation does not change node cardinalities but can reduce tuple width at the selected node and its ancestors. 
When the output of a pushed-down computation is smaller than its input, subsequent joins transfer and process fewer features, thereby lowering I/O cost. In addition, when multiple sub-computations are pushed down to the same node, aggregating their outputs may further decrease the tuple width.
The join's CPU cost remains unaffected since tuple counts are unchanged, but reduced tuple width improves data movement efficiency across the plan.

}}

%


\section{ Problem Formalization}
\label{sec:formalization}

In this section, we formalize the problem of optimizing the factorization and push-down plan of inference computations over an $m$-way join following a specific order.  
We then demonstrate that the above optimization problem can be decoupled from the join order optimization without compromising optimality.

\vspace{-10pt}
\subsection{\textcolor{black}{Naive} Factorization Optimization Problem}
\label{sec:factorize-optimize-prob}

Without loss of generality, an $m$-way join query, following a specific optimized join order with all predicates and projections pushed down, is viewed as a binary tree consisting of $2m-1$ nodes, a common abstraction in database query optimizers~\cite{rogov2023postgresql, garcia2008database}. Each of the $m$ leaf nodes $N_1,...,N_m$ represents a \texttt{scan} operator that outputs all tuples from a \textit{source table} $R_i$ to be joined. In addition, each of $m-1$ non-leaf nodes, $N_{m+1},...,N_{2m-2}$, stands for an intermediate \texttt{join} node. The \texttt{join} operator at the root node (denoted as $N_{2m-1}$) of the tree outputs the final join result, which will be consumed by a factorizable computation $f$.
$f$ may range from a simple linear algebra operator, such as a left matrix multiplication operator, to an arbitrary inference application, e.g., the model in Fig.~\ref{fig:user-defined-workflow}. $f$ could be factorized into $m$ disjoint sub-computations $f_1, ..., f_{m}$ so that $f_i$ only depends on the data associated with the leaf node $N_i$, $i=1,\dots,m$,  and we have $f(\boldsymbol{x})=aggregate(f_1(\boldsymbol{R_1}), ..., f_{m}(\boldsymbol{R_{m}}))$, where $\boldsymbol{x}=\boldsymbol{R_1}\bowtie...\bowtie\boldsymbol{R_{m}})$.  $f_i$ is only eligible to be pushed down to node $N_i$ or each ancestor node of $N_i$. Pushing down $f_i$ to the root node $N_{2m-1}$ means $f_i$ has not been pushed down.

The factorization optimization problem is an assignment problem~\cite{ramshaw2012minimum} by assigning $m$ factorized computations to $2m-1$ nodes to optimize the overall performance by minimizing a global cost function $c(\boldsymbol{x})$ that evaluates the benefit of the assignment $\boldsymbol{x}=\{x_{ij}\}$, $i=1,\dots,m$, $j=1,\dots,2m-1$. If $f_i$ is pushed down to node $N_j$, $x_{ij}=1$, otherwise, $x_{ij}=0$. 

\begin{figure}[ht]
\centering
\includegraphics[width=0.48\textwidth]{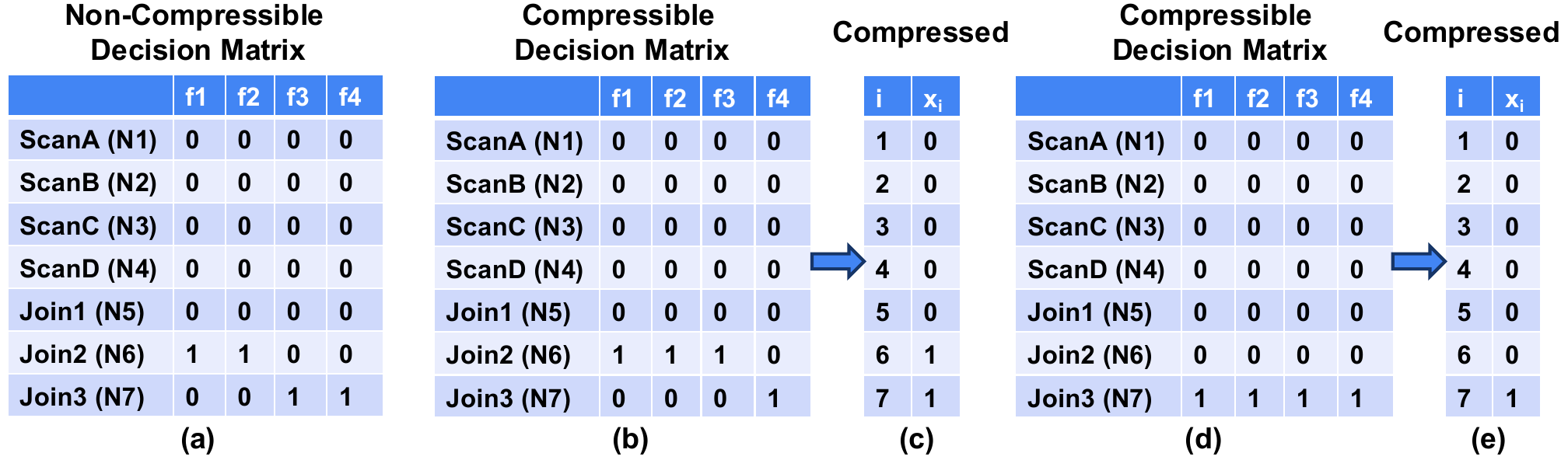}
\caption{\small Compressible and Non-Compressible Decision Matrices}
\label{fig:compressed-decision}
\end{figure}

Taking the query plan in Fig.~\ref{fig:example-1} as an example, 
its push-down decision matrix to optimize is shown in Fig.~\ref{fig:compressed-decision}(a), where $x_{ij}$ indicates whether the subcomputation $f_j$ should be pushed to node $N_i$.

\begin{figure}[h]
\centering{%
   \includegraphics[width=3in]{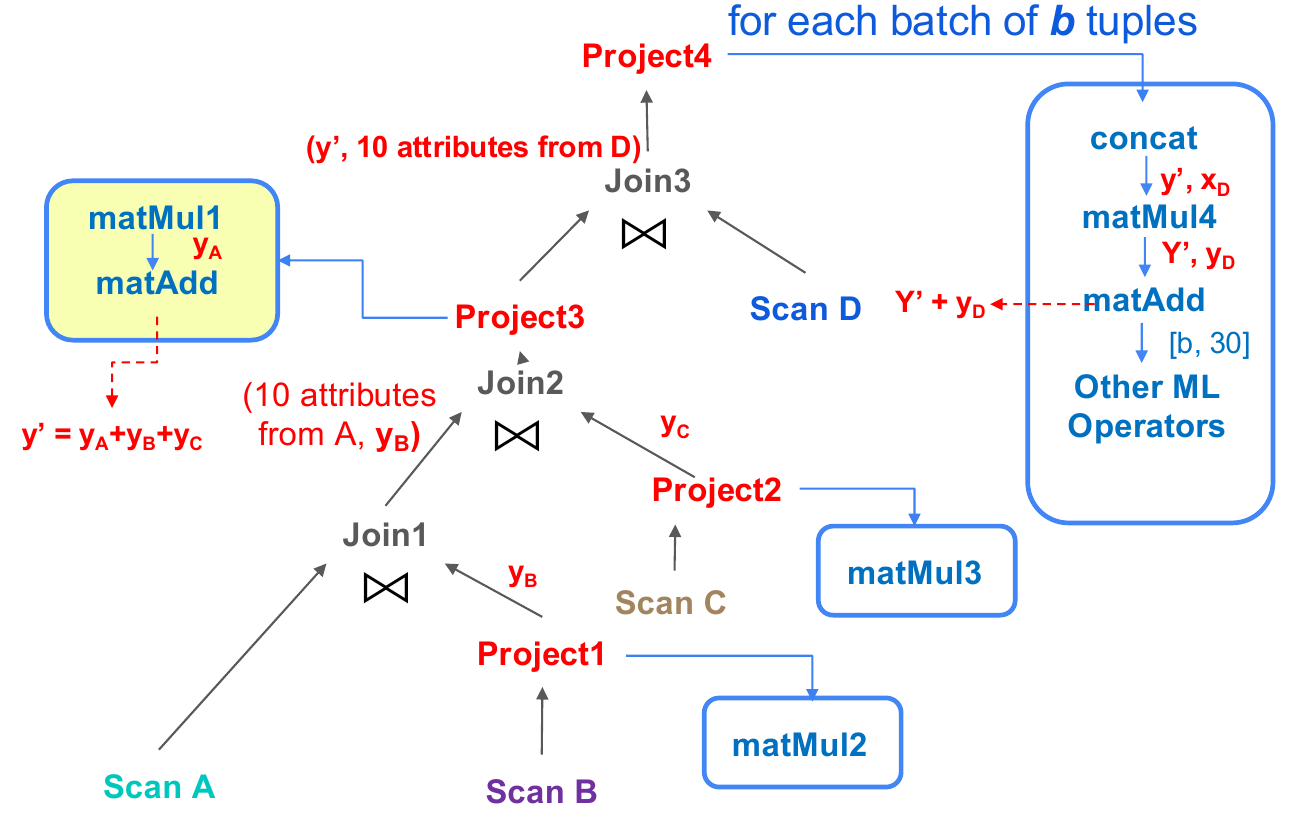}  
}
\caption{\label{fig:example-2} \small {Aggressive Aggregation on the Example in Fig.~\ref{fig:example-1}}}

\end{figure}

\vspace{-20pt}
\textcolor{black}{\subsection{New Problem Formalization}}
\textcolor{black}{However, the naive problem did not consider} that when multiple factorized computations are pushed to a node---or when the results of such computations are propagated to it---it is possible to further optimize the output size at that node by \textit{aggressively aggregating} the results of all the incoming factorized computations. \textcolor{black}{For example, in Fig.~\ref{fig:example-2}, at the expression for node \textbf{Project3} in the rewritten query plan (after applying optimized push-down decisions), a \texttt{matAdd} operator aggregates the outputs of descendants, $y_A$, $y_B$, and $y_C$, all are $30$-dimensional vectors.} This aggregation reduces the output size from $30 + 30 + 30 = 90$ to just 30 dimensions per tuple. 
%
%
\textcolor{black}{This observation motivates us to \textit{group} factorized computations and push them jointly to a node whenever possible. 
At each node, we consider only those computations that are \textbf{eligible for push-down} and \textbf{not yet assigned to any descendant node}. 
Once a computation has been pushed to a subtree, it is \textbf{consumed and no longer available}  at ancestor nodes. 
We then choose between two options: (i) perform no push-down at this node, or (ii) push down \textit{all eligible factorized computations} to fully exploit aggregation opportunities. 
We refer to this as a \textbf{grouped push-down strategy}, which naturally excludes non-pushable operators while ensuring that pushed computations are not redundantly applied higher in the join tree.}


For example, Fig.~\ref{fig:compressed-decision} lists three assignment plans for the query given in Fig.~\ref{fig:example-1}. The plan in Fig.~\ref{fig:compressed-decision}(a) only pushes down $f_1$ (\texttt{matMul1}) and $f_2$ (\texttt{matMul2}) to $N_6$ (\texttt{Join2}), while $f_3$, which has not been pushed to any descendant nodes of $N_6$ and is thus available at $N_6$, is not pushed to $N_6$, but pushed to $N_7$.  
Therefor, it does not adhere to the \textcolor{black}{grouped push-down strategy}. It should be pruned for the following reasons. If $N_6$ has a lower output cardinality than $N_7$, the plan in Fig.~\ref{fig:compressed-decision}(b) will outperform both plans in Fig.~\ref{fig:compressed-decision}(a) and in Fig.~\ref{fig:compressed-decision}(d), since pushing all $f_1$, $f_2$, and $f_3$ to $N_6$ will save more computational overhead (due to lower cardinality) and will also save more joining cost (due to the size reduction brought about by aggressive aggregation at $N_6$). Otherwise, if $N_7$ has a lower output cardinality than $N_6$, the plan in Fig.~\ref{fig:compressed-decision}(d) outperforms both plans in Fig.~\ref{fig:compressed-decision}(a) and Fig.~\ref{fig:compressed-decision}(b) for similar reasons.
Both plans in Fig.~\ref{fig:compressed-decision}(b) and Fig.~\ref{fig:compressed-decision}(d) follow the \textcolor{black}{grouped push-down strategy}. 

Since we only consider the plans following \textcolor{black}{grouped push-down strategy}, the decision matrices could be compressed into vectors $\boldsymbol{x}=\{x_i\}$, $i = 1,...,2m-1$, as illustrated in Fig.~\ref{fig:compressed-decision}(c) and (e), where $x_i$ represents whether to push down all available factorized computations at $N_i$ \textcolor{black}{, denoted as $D_i$, to $N_i$ itself}. 
{\color{black}
Then the cost corresponding to a decision vector $\boldsymbol{x}$ is denoted as $b(\boldsymbol{x})$ as shown in Eq.~\ref{eq:cost-function}. 
   \begin{equation}
    \label{eq:cost-function}
    \color{black}
    \small
     \begin{split}
    b(\boldsymbol{x}) 
    &= inference\_computation\_cost + \lambda \times join\_io\_cost \\
   &= 
    \sum_{i:\,x_i=1} n_i \cdot u(x_{\prec i})+\lambda \cdot \sum_{i=1}^{2m-1} 
             n_i\cdot w(x_{\preceq i})
             \end{split}
    \end{equation}
The first term $\sum\limits_{i:\,x_i=1} n_i \times u(x_{\prec i})$ represents the cost of performing the available factorized computations at $N_i$. Here, $n_i$ is the number of tuples emitted at $N_i$, and 
$u(x_{\prec i})$ is the computational cost per tuple at node $N_i$, which depends on previous decisions $x_{\prec i}$. It can be expanded to $u(x_{\prec i})=u(\{f_j| f_j \in D_i\})$. Note that $D_i$ represents all factorized sub-computations that satisfy: (1) they correspond to the leaf nodes of the subtree ($T_i$) rooted at $N_i$, and (2) they haven't been pushed to any descendant nodes of $N_i$, based on $x_{\prec i}$, i.e., $D_i=\{f_j | N_j \in T_i.leaves \wedge (\nexists N_k \prec N_i | f_j\in D_k \wedge x_k=1) \}$. Here, $N_j \prec N_i$ means $N_j$ is a descendant node of $N_i$.
Nodes with $x_i=0$ do not have computations pushed down and will not contribute to the inference computation cost, so they are excluded.

In the second term, $w(x_{\preceq i})$ is the tuple width at $N_i$, which depends on $x_i$, the width reduction ratio of $f_j\in D_i$, and the input sizes. The input sizes are determined by the prior push-down decisions $x_{\prec i}$, since width reductions at one node propagate upward through its ancestors until reaching the root or a node that performs its own push-down. The high-level structure of this I/O term is derived from the existing I/O cost estimation for two-way joins ~\cite{shapiro1986join, kumar2015learning}.

Finally, we omit the CPU cost of joins, since, as explained in Sec.~\ref{sec:benefits}, 
factorization and push-down do not affect node cardinalities, making this cost constant.


Therefore, our goal is to minimize  $b(\boldsymbol{x})$ as formalized in Eq.~\ref{eq:new-objective}.}
    %
    %
    Eq.~\ref{eq:new-constraint2} defines that if $D_i$, the set of all available factorized computations at $N_i$, is empty (pushed to descendant nodes of $N_i$), $x_i=0$. \textcolor{black}{Eq.~\ref{eq:new-constraint3} indicates that $D_{2m-1}$, which is the set of factorized computations that have not been pushed to any non-root tree node, and thus available at the root node $N_{2m-1}$, must be computed at $N_{2m-1}$, i.e., $x_{2m-1}=1$.}
    \begin{equation}
    \label{eq:new-objective}
    \color{black}
    \small
     \vspace{-15pt}
    \begin{split}
    \underset{\boldsymbol{x}\in\{0,1\}^{2m-1}}{\operatorname*{argmin}}\;
    \sum_{i:\,x_i=1} n_i \cdot u(x_{\prec i})+\lambda \cdot \sum_{i=1}^{2m-1} 
             n_i\cdot w(x_{\preceq i})
    \end{split}
    \end{equation}

    \begin{equation}
    \label{eq:new-constraint2}
    \small
    \vspace{-15pt}
    x_{i} = 0, \text{ if } D_i = \textcolor{black}{\emptyset}
    \end{equation}
    
    \begin{equation}
    \label{eq:new-constraint3}
    \small
    \color{black}{x_{2m-1} = 1, \text{ if } D_{2m-1} \neq \emptyset}
    \end{equation}
    
    \vspace{-4pt}
    {\color{black}{\subsection{Why Naive Solutions Cannot Work?}
    
    Next, we explain why the below solutions cannot work well.

    \noindent
    \textbf{No Push-down.}
$S_{original}=n_{2m-1}\times u_f$ denotes the overall inference computation cost without factorization and push-down, where $n_{2m-1}$ is the number of tuples emitted from the root node of the join tree, and $u_f$ represents the computational cost of the factorizable operator $f$ (before factorization).
%
Since $u_f$ equals to $\sum_{i:\,x_i=1} u_i(x_{\prec i})$ (after factorization) in Eq.~\ref{eq:new-objective}, if there exists a non-root node $N_j$ with smaller cardinality ($n_j < n_{2m-1}$) and/or reduced output width ($w_j - w'_j > 0$), setting $x_j = 1$ yields a lower cost than the no–push-down plan ($x_{2m-1}=1$, $x_i=0$ for $i<2m-1$).

    \noindent
    \textbf{All Push-down.} 
This strategy pushes each factorized computation directly to its base table, leaving no computation at intermediate nodes. 
However, as implied by Eq.~\ref{eq:new-constraint3}, if an intermediate or root node $N_j$ (for $m < j \leq 2m-1$) is an ancestor of table node $N_i$ ($1 \leq i \leq m$) and has smaller cardinality ($n_j$<$n_i$) with smaller or same width reduction ratio, pushing $f_i$ to $N_j$ can yield a lower cost than pushing it to $N_i$.

\noindent
\textbf{Sorting and Caching}
    Another approach is to materialize and sort the join output, then apply inference with caching to avoid redundant computations. 
However, this method cannot leverage the I/O reductions achieved by (partial) factorization and push-down (the second term in Eq.~\ref{eq:new-objective}), and thus cannot ensure global optimality. 
In addition, the cache size is bounded by the available memory and it introduces additional sorting and lookup overhead. 
Nevertheless, caching can complement our method when the optimal push-down plan cannot fully eliminate redundancy (e.g., when pushing a computation to an intermediate join node that has lower cardinality than base tables, but higher redundancy). In addition, performing factorization first can further reduce cache size and lookup cost.

%

\noindent
\textbf{Extending Decision Rules for Two-Way Joins}
%
Morpheus~\cite{factorize-la} decides whether to factorize and push down computations for a two-way join between $R$ and $S$ using a simple rule: $n_S/n_R > 5$ or $w_R/w_S > 1$, where $n$ denotes tuple count and $w$ denotes attribute count. 
However, this rule ignores key factors such as the cost of each factorized computation, tuple width reduction, and intermediate join cardinalities, limiting its applicability to multi-way joins.

\noindent
\textbf{A Bottom-Up Traversal}
%
A simple bottom-up strategy minimizes the output size at each join node. 
However, this is suboptimal because reductions in redundant inference (first term in Eq.~\ref{eq:new-objective}) and I/O cost (second term) at node $N_i$ from pushing down $f_j$ depend on multiple factors: 
$f_j$’s feature elimination capability, its computational complexity, and the output cardinality of $N_i$. 
It also overlooks the opportunity cost that computations pushed to lower nodes become unavailable to their ancestors.

    }
    
    }

\vspace{-14pt}
\textcolor{black}{\subsection{NP-Hardness Discussion and Formal Proof}}
\label{sec:np-hard}

{\color{black}

The push-down planning problem, defined by Eqs.~\ref{eq:new-objective}, \ref{eq:new-constraint2}, and \ref{eq:new-constraint3} 
 is NP-hard, since it can reduce from the \emph{Tree-like Weighted Set Cover (TWSC)} problem~\cite{guo2006exact},
for which, the decision problem is NP-Complete, while the optimization problem is NP-hard. 

\noindent
\textbf{Tree-like Weighted Set Cover (TWSC)} We are given a base set $S=\{s_1, \dots, s_p\}$ and a tree-like collection $C$ of subsets of $S$, $C=\{c_1,\dots,c_q\}$, $c_i \subseteq S$ for $1\leq i \leq q$, with $\bigcup\limits_{1\leq i \leq q} c_i=S$. In addition, the subsets in $C$ can be organized in a tree $T$, such that every subset one-to-one corresponds to a node of $T$, and every node is a subset of its parent node. Each subset in $C$ has a positive real weight $v(c_i)>0$ for $1\leq i \leq q$. The weight of a subset collection is the sum of the weights of all subsets in it. 

The problem is to find a subset $C'$ of $C$ with minimum weight, which covers all elements in $S$, i.e., $\bigcup\limits_{c\in C'}c=S$. 

\noindent
\textbf{Reduction.}
From any TWSC instance, we can construct a special instance of our problem, as follows.

\begin{enumerate}[leftmargin=10pt,itemsep=2pt,topsep=2pt]
\item \textbf{Join tree.} We transform $T$ into a new binary tree $T'$, so that every node in $T$ is a node in $T'$. We ensure that a node in $T'$ is exactly the union of elements in its children nodes, by adding new leaf nodes. We also add intermediate neutral nodes to maintain the binary tree structures, when needed.

For example, given 
$S=\{s_1, s_2, s_3, s_4\}$ and
$c_1=\{s_1, s_2, s_3, s_4\}$, $c_2=\{s_2, s_3\}$, and $c_3=\{s_1, s_3\}$, we will add $c_4=\{s_4\}$ and a node $c_5=\{s_2, s_3, s_4\}$, so that $c_2$ and $c_4$ are children nodes of $c_5$, while $c_3$ and $c_5$ are children nodes of $c_1$, forming an expanded set $C^*=C\bigcup \{c_4, c_5\}$. 

In this way, the union of elements at all leaf nodes will be $S$, and each leaf node represents a factorized subcomputation. 
For each newly added node, we set its weight to a sufficiently large number $R >> max(v(c_i))$ so that it will not be selected for push-down in the optimal plan. (For each node in $T'$, we set its cardinality $n_i=1$.) 

\item \textbf{Inference Computation Costs.} For each node $N_i\in T'$, it corresponds to a node in $c_i \in C^*$, we set 
$u(x_{\prec i}) := v(c_i)$.

\item \textbf{I/O Costs.} We set $\lambda=0$ so that the I/O term is removed.
\end{enumerate}

\noindent
\textbf{Objective equivalence.}
Under the above construction, we have
\begin{equation}
\small
b(\boldsymbol{x})
= \sum_{i:\,x_i=1} (v
(c_i) )
\label{eq:convert}
\end{equation}
Guaranteed by Eq.~\ref{eq:new-constraint3} and the large penalties brought by newly added nodes, the factorized sub-computation at each leaf node will be pushed to a selected node in $T$ corresponding to $c_i \in C$, with $x_i$ set to $1$. Let $x_i=1$ indicate $c_i$ is selected, i.e., $c_i\in C'$. Then we must have $\bigcup\limits_{c\in C'}c=S$, because the computation at each leaf node must be assigned to some ancestor nodes that contain all elements at the leaf node, given the subset relationship between parent and child in the underlying tree structure.
Therefore, minimizing $b(\boldsymbol{x})$ is equivalent to minimizing the total weights of all selected subsets in $C'$. Hence, an optimal solution of our problem corresponds exactly to an optimal TWSC solution.

The transformation is polynomial in $p+q$ and preserves optimality; therefore, our problem is NP-hard.
\hfill$\square$
}

\vspace{-3pt}
\noindent
\textbf{Search Space Analysis:} If we naively apply the exhaustive search to resolve the above problem, the time complexity is exponential due to the huge search space. For example, if the given $m$-way join follows the left-deep join order, supposing the total number of available factorization plans is $f(m)$, we have $f(2)=4$ (i.e., push-down at left leaf and root, push-down at right leaf and root, push-down to left and right leaves, push-down only at root), and when $m>2$, $f(m) = (f(m-1)+3)\times 2$. That's because (1) the left subtree of the new root node has $f(m-1)+3$ factorization plans (the old root node now becomes an intermediate node, bringing three new options), and (2) the right subtree only consists of one node, thus having two factorization plans (i.e., push down or not). Therefore, $f(m) =5\times 2^{m-1}-6$. Most of the other join orders lead to roughly similar search space sizes.

\vspace{-10pt}
\subsection{Join Order vs. Factorization Optimization}
\label{sec:decoupled-join-factorize}
We propose first optimizing the join order and then optimizing the factorization and push-down plan for below reasons.

\noindent
\textbf{1. Decoupled optimization leads to search space reduction:} Given an $m$-way join, combining the optimization of join order and factorization will lead to a huge search space of $m!\times (5\times2^{m-1} - 6)$. 
Decoupling the two steps will significantly reduce the search space.

\vspace{3pt}
\noindent
\textbf{2. Decoupling join order optimization and factorization optimization has minimal impact on optimality:} 
Given a join tree following the optimal ordering $T$, supposing the factorization optimization process identifies an optimal factorization plan denoted as $optFactorize(T)$. Then, we argue that given a join tree with sub-optimal ordering $T'$, its optimal factorization plan, denoted as $optFactorize(T')$, will probably achieve worse performance than $optFactorize(T)$. 
That's because an optimal ordering usually reduces the cardinalities (i.e., determining the CPU cost) and the sizes (i.e., $size = cardinality$ $\times tupleSize$, determining the memory and I/O costs) of the outputs at intermediate join nodes, compared to sub-optimal orderings. 
Given any factorized computation $f_j$ which is pushed to node $N_i$ of join tree $T$ in $optFactorize(T)$, and pushed to node $N'_k$ in join tree $T'$ in $optFactorize(T')$, the cardinality and output size at $N_i$ in $T$ is more likely to be smaller than that at $N'_k$ in $T'$. Therefore,  the overall cost $optFactorize(T)$ is more likely to be smaller than $optFactorize(T')$, according to Eq.~\ref{eq:new-objective}.
Our empirical results are consistent with the above analysis as shown in \textcolor{black}{Sec.~\ref{sec:factors-analysis}}. 

\vspace{5pt}
\noindent
\textbf{3. Decoupled optimization facilitates the reuse of existing query optimizers:} Existing join ordering optimizers could be easily dropped into our system with minimal integration effort.

\section{Factorization Optimization}
\label{sec:optimization}

To solve the optimization problem, we propose an iterative \textit{Genetic} algorithm and a \textit{Greedy} algorithm.

\subsection{Genetic Optimizer}
\label{sec:genetic}

\begin{figure*}[hbt]
\centering
\includegraphics[width=0.96\textwidth]{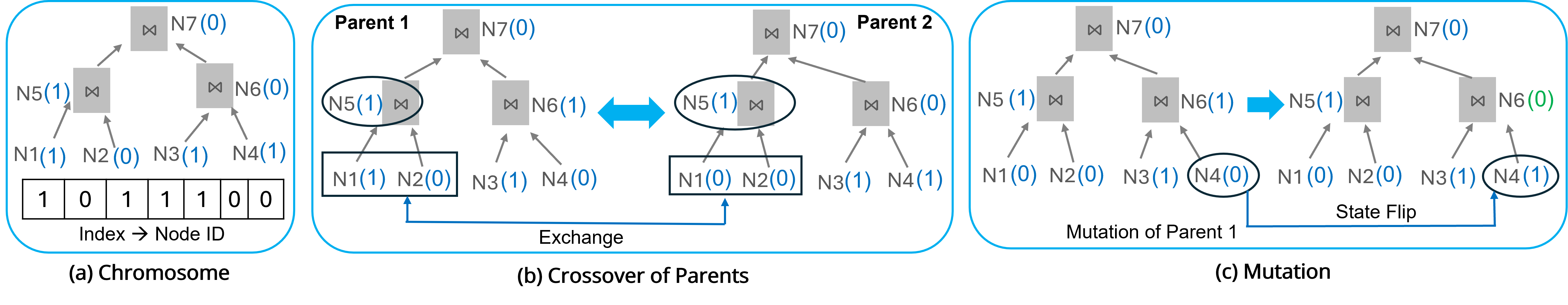}
\vspace{-12pt}
\caption{\small Illustration of Genetic Optimization}
\label{fig:genetic-optimizer}
\vspace{-6pt}
\end{figure*}

We first propose a novel application of the genetic algorithm~\cite{steinbrunn1997heuristic} to iteratively enumerate and evaluate promising factorization plans. 
Similar to classical genetic algorithms~\cite{steinbrunn1997heuristic}, it includes the following components. (The key steps of enumerating new promising candidate plans at each generation are illustrated in Fig. \ref{fig:genetic-optimizer}.)

\noindent
\textbf{Chromosome Encoding:} A chromosome denotes a factorization plan for a given join order tree, represented as an array. Each entry in the array corresponds to a node in the join tree and specifies whether we shall push all available factorized computations at that node (See Fig.~\ref{fig:genetic-optimizer}(a)). 

\noindent
\textbf{Population Initialization:} The genetic optimization process requires first initializing $m$ factorization plans (i.e., chromosomes) to form an initial population (lines 1-5 in Alg.~\ref{alg:genetic-optimization}). We initialize a chromosome by randomly setting the states of all nodes to $0$ or $1$. However, a chromosome becomes invalid if a parent node is in state $1$ but none of its child nodes is in state $0$. Therefore, we need to look for invalid states in the initialized chromosomes and update them to \textcolor{black}{$0$} if such cases are found (line 4). 

\noindent
\textbf{Generations:} The genetic optimization algorithm will run $g_{max}$ iterations (i.e., generations). In each generation (lines 6-12), it performs crossover and mutation to generate $k$ new chromosomes.  For crossover, we randomly select two parent chromosomes from previously initialized chromosomes and perform a crossover between the parents. As depicted in Fig.~\ref{fig:genetic-optimizer}(b), a non-leaf node is randomly picked as a crossover point (e.g., node N5 in Fig.~\ref{fig:genetic-optimizer}(b)), and the subtree rooted at the picked node is exchanged between the two selected parents.
The result of crossover to parent-1 is the left chromosome in Fig.~\ref{fig:genetic-optimizer}(c). During the mutation, we randomly select a node and flip its state ($0\leftrightarrow1$). Similar to line 4, both crossover and mutation functions contain a step of fixing invalid states. 
For example, as shown in Fig. \ref{fig:genetic-optimizer} (c), after mutating node N4, the state of N6 becomes $1$ while its child nodes are also $1$, and it was fixed by changing the state to $0$. After generating $k$ new chromosomes via crossover and mutation, the fitness of all chromosomes in the population will be evaluated, and $m$ chromosomes with the best fitness will be selected as the next generation of the population. 
Higher values of $g_{max}$, $m$, and $k$, will explore a larger number of factorization plans with increasing optimization latency. On the other hand, lower values will make the optimizer run faster while increasing the possibility of missing the optimal plan.

\begin{algorithm}[h]
\small
    \SetKwInOut{getSources}{Objective}
    \getSources{Genetic Optimization\\}
    \ResetInOut{Output}
    \SetKwInOut{Input}{Input}
    \SetKwInOut{Output}{Output}
    \Input{$nodes$: a list of nodes in the binary join tree; $g_{max}$: maximum generation count; $m$: chromosome population size; $k$: number of new chromosome to generate at each generation }
    \Output{Chromosome with highest utility value}

$count$ $\gets$ $0$; $P$ $\gets$ $\Phi$ \;
\While{$count < m$}{
$p$ $\gets$ $array(random(0, 1)$ $for$ $i \in 1, 2, .., |nodes|)$ \;
set any node whose child nodes with state $1$ back to \textcolor{black}{$0$}\;
$P$ $\gets$ $P \cup \{p\}$;
$count++$\;
}

\For{$i=1,...,g_{max}$}{
\While{$|P| < m+k$}{
$p1$, $p2$ $\gets$ randomly selected chromosomes \;
$p'1$, $p'2$ $\gets$ $crossover(p1, p2)$\;
$p1$ $\gets$ $mutate(p1)$ ;
$p2$ $\gets$ $mutate(p2)$ \;
$P$ $\gets$ $P \cup \{p_1, p_2\}$\;
}
$P$ $\gets$ the set of top-$m$ plans in $P$ that has the best fitness\;
}

\textbf{return} the top plan in $P$ that has the best fitness;
    \caption{\small Genetic Optimization}
    \label{alg:genetic-optimization}
\end{algorithm}

\noindent
\textbf{Fitness Evaluation:}  Lastly, we estimate the fitness of a chromosome using the optimization objective function in \textcolor{black}{Eq.~\ref{eq:new-objective}}. 
We estimate the cost of all factorized computations ($w_i$) using a neural-symbolic approach. We first use linear regression models to learn the cost ratio of each ML operator (e.g., matrix\_multiply and other linear algebra operators, decision forest, distance computation, etc.) in terms of the input feature size and model dimensions. Then, as detailed in Sec.~\ref{sec:ir}, each factorized computation is abstracted as an analyzable expression, with its logic described as a computational graph, with each node being an ML operator (e.g., concat, matrix multiplication) or an expression operator (i.e., arithmetic, relation, bit, conditional operators). Then the overall cost of an expression is estimated by traversing the computational graph. 

\subsection{Greedy Optimizer}
\label{sec:greedy}

The efficiency and accuracy of the Genetic algorithm highly depend on the tuning of hyper-parameters such as the number of generations ($g_{max}$), the size of the population ($m$), and the number of new chromosomes at each generation ($k$). In addition, the evaluation of the entire plan tends to be expensive and less accurate for complicated factorizable AI/ML computations. To address the limitation, we propose a greedy optimizer following the ``Benefit-First'' idea, which dynamically estimates cost reduction at each node and selects nodes that are expected to yield the highest positive cost reduction if a push-down is applied. By prioritizing high-impact nodes, it efficiently explores the join tree in linear time. It forms an optimal factorization plan while avoiding the prohibitive cost of exploring the full exponential search space. 

\vspace{2pt}
\noindent
\textbf{Our Benefit-First Algorithm} is formalized in Alg.~\ref{alg:greedy-optimization}. The state of every node in a multi-way join tree (with join ordering fully optimized and predicate push-down applied) is set to 0 initially, representing that no factorized computation will be pushed down. We iterate over nodes (with cost $< cost\_threshold$) in the ascending order of their cost, utilizing a priority queue, and decide to iteratively factorize the node with the minimum cost by changing the state to $1$. Before changing the state to $1$, we check if all child nodes have already been factorized and change the state to $2$ if such cases are found (lines 9-10). These states will be changed back from $2$ to $0$ at the end (line 18). 
For a node $N_i$ selected for push-down, since we will push all available computations ($D_i$) at $N_i$, no more computations from $D_i$ will be pushed to its descendant nodes. Note that any descendant node $N_j$ of $N_i$ has $D_j \subseteq D_i$. It means we will never change the states of $N_j$'s descendant nodes, so we mark all of its non-processed descendants as processed (line 17). 
In addition, the decision to push down all factorized computations at a node changes its ancestor nodes' available factorized computations. So, each time a node is selected, we update the cost of the ancestor nodes and adjust the priority queue accordingly (lines 12-16). 

\vspace{-2pt}
\begin{algorithm}[h]
    \small
    \SetKwInOut{getSources}{Objective}
    \getSources{Greedy Algorithm for Pushing Down Operators \\}
    \ResetInOut{Output}
    \SetKwInOut{Input}{Input}
    \SetKwInOut{Output}{Output}
    \Input{$nodes$: a list of all nodes in the join tree; $t$: cost threshold }

$pq$ $\gets$ $PriorityQueue()$ \;
\ForEach{$node \in nodes$}{
$node.state$ $\gets$ $0$ \;
$node.cost$ $\gets$ $getCost(node)$ \;
\If{$node.cost < t$}{
push $(node.cost, node)$ to $pq$ \;
}
}

\While{$pq$}{
$curCost$, $node$ $\gets$ $pq.get()$ \;
\If{$node.leftChild.state \neq 0 \land node.rightChild.state \neq 0$}{\tcp{No available factorized computations at $node$ }
$node.state$ $\gets$ $2$ \;
}
\Else{
$node.state$ $\gets$ $1$ \;
\tcp{update the cost of ancestor nodes of $node$ in $pq$}
$curNode$ $\gets$ $node$ \;
\While{$curNode.parent \neq NULL$}{
 $updateCost(curNode.parent, pq)$ \;
 $curNode$ $\gets$ $curNode.parent$ \;
}
}

remove $node$ and all its descendant nodes from $pq$\;
}
reset all nodes with $state=2$ to $state=0$; \textbf{return}\;

    \caption{\small Greedy Optimization}
    \label{alg:greedy-optimization}
\end{algorithm}

\noindent
\textbf{Cost Estimation:} We use Sobol's method to estimate the per-node cost by identifying factors determining cost reduction associated with pushing all available factorized computations to the node. 

Sobol's method~\cite{sobol-1, sobol-2} is a technique to analyze the sensitivity of a (black-box) function to its variables based on a variance analysis methodology. Assume that the output of the model $Y$ is a function $f(X)$ where $X$ is a set of $d$ input variables $\{X_1, X_2, ..., X_d\}$ independently and uniformly distributed such that $X_i \in X \in [0, 1]$. The variance of $Y$ is calculated according to Equation \ref{eq:variance-y}, where the variance of $X_i$, denoted as $Var(X_i)$, is the variance of $E(Y|X_i)$. Here, $Var(X_{ij})$ is defined as $Var(E(Y|X_iX_j)) - Var(X_i) - Var(X_j)$. The Sobol's sensitivity index is derived using the formula $S(X_i) = Var(X_i) / Var(Y)$. The sensitivity index for the first order and higher order interactions of inputs should follow Eq.\ref{eq:sobol-index}.

\begin{equation}
\small
        \vspace{-8pt}
    Var(Y) = \sum_{i=1}^d Var(X_i) + \sum_{i < j}^d Var(X_{ij}) + ... + Var(X_{12...d})
    \label{eq:variance-y}
\end{equation}

\begin{equation}
\small
    \vspace{-3pt}
    \sum_{i=1}^d S(X_i) + \sum_{i < j}^d S(X_{ij}) + ... + S(X_{12...d}) = 1
    \label{eq:sobol-index}
\end{equation}

We first identify factors that determine the cost reduction (or benefits) of pushing down available factorized computations to a node in the join tree, and then calculate Sobol's sensitivity index for the first-order and second-order interaction of these factors.

%
%
Examples of results are shown in Tab.~\ref{tab:sobol-sensitivity}. 
Those interactions not listed in the table have negligible sensitivity indices.  
We then apply logistic regression (on the first order and second order interactions of factors listed in Tab.~\ref{tab:sobol-sensitivity}) to learn a cost estimator. 

\vspace{-5pt}
\begin{table}[h]
\caption{\small  Important Cost Factors Identified by Sobol's Method Evaluated Using $50$ Representative Synthetic Queries on Datasets Described in Sec.~\ref{sec:datasets}. Here, S. Index refers to the sensitivity index.}
\label{tab:sobol-sensitivity}
\vspace{-8pt}
{
\scriptsize
\begin{tabular}{|p{6.7cm}|p{1cm}|}
\hline
 Factors & S. Index \\ \hline
 Cardinality Ratio (i.e., the number of tuples output from the node divided by the number of tuples input to the node)   & 0.325 \\ \hline
 Factorized Computations Cost (i.e., the cost of all available factorizable computations (See ``Fitness Evaluation'' of Sec.~\ref{sec:genetic})) & 0.216 \\\hline
Tuple Dim Ratio  (i.e., the ratio of the size of each tuple after pushing down all available factorized computation at this node to the size of each tuple before pushing down)   & 0.131         \\ \hline
 Depth Ratio (i.e., the level of node divided by tree height)    & 0.070 \\ \hline
 Cardinality Ratio,  Factorized Computations Cost & 0.068  \\ \hline
 Cardinality Ratio, Tuple Dim Ratio & 0.064  \\ \hline
Cardinality Ratio, Depth Ratio & 0.052  \\ \hline
\end{tabular}
}

\end{table}

\section{Our Factorization Framework}
\label{sec:framework}

In this section, we will introduce the design and implementation of the end-to-end framework for analyzing the inference queries, optimizing, and applying factorization plans.

\subsection{Factorizable IR}
\label{sec:ir}
We enable factorization optimization by representing an arbitrary inference workflow as an analyzable intermediate representation (IR), customizing relational algebra with expressions. Each expression is lowered to a computation graph for factorization analysis and other optimization analysis.
As illustrated in the left part of Fig.~\ref{fig:user-defined-workflow}, the query logic is represented in relational algebra. A relation $R$ is a collection of tuples $t$, following the schema $(A_1,...A_k)$, i.e., $R(A_1,...A_k)=\{t\}$. Each node represents a relational operator such as $join$ and $aggregate$. Some relational operators, such as $filter$ (i.e., $select$) and $project$ (i.e., similar to $map$ or $transform$), can be customized using expressions as follows:

\noindent
(1) A $filter$ operator filters the tuples in a relation, based on a predicate or an expression $e_{fil}$ that returns a boolean value. 

\begin{DSL}
filter$\{e_{fil}$: $(A_1,...A_k)$ $\rightarrow$ bool$\} (R)=\{t | t \in R \wedge e_{fil}(t) = true\}$
\end{DSL}

\noindent
(2) A $project$ operator transforms the input relation $R$ by filtering attributes or applying an expression $e_{proj}$ to transform each tuple $t\in R$ into a tuple with a different schema (for example, concatenating multiple attributes of type \texttt{Float} into a single attribute of type \texttt{Array<Float>}).

\begin{DSL}
project$\{e_{proj}$: $(A_1,...A_k)$ $\rightarrow$ $(A'_1,...A'_l)$$\}(R)=\{e_{proj}(t) | t \in R\}\}$
\end{DSL}

\vspace{3pt}
\noindent
As shown in the right part of Fig.~\ref{fig:user-defined-workflow}, expressions used to customize the above relational operators, such as $e_{fil}$ and $e_{proj}$, are represented as a lower-level IR, similar to a computational graph. In the graph, each node is a built-in ML operator, an opaque UDF, or a higher-order expression operator. Each edge represents a dataflow from the source node to the target node. We currently support built-in atomic operators such as \textit{matMul} (matrix multiplication), \textit{matAdd} (matrix addition), \textit{relu}, \textit{Softmax},  \textit{dt} (decision tree), \textit{xgboost},
\textit{nnSearch} (nearest neighbor search), \textit{conv2D},
\textit{batchNorm},
\textit{concat}, \textit{binarization}, 
\textit{standardScaler}, and \textit{minMaxSca}\textit{ler}. The linear algebra functions rely on the Eigen~\cite{eigenweb} and LibTorch~\cite{libtorch_cuda} libraries. XGBoost is implemented using Google's Yggdrasil library~\cite{guillame2022yggdrasil, tf-df, guan2023comparison}. Other functions are hand-coded from scratch following scikit-learn implementations~\cite{scikit-learn}. 

Then, a set of higher-order functions are provided to compose these
atomic functions/operators into a new expression that can be regarded as a
directed acyclic graph of functions, which includes:

\noindent
(1) The boolean comparison operations: \texttt{==}, \texttt{>}, \texttt{!=}, etc.;

\noindent
(2) The boolean
operations: \texttt{\&\&}, \texttt{||}, \texttt{!}, etc.;

\noindent
(3) The arithmetic operations: \texttt{+}, \texttt{-}, \texttt{*}, etc.;  

\noindent
(4) conditional branch: \texttt{$l_1$?$l_2$:$l_3$}, or \texttt{if($l_1$) $l_2$ else $l_3$}.

\noindent
(5) function call: e.g., f1(f2(x)). 

\vspace{-8pt}
\subsection{Factorization Analysis and Implementation}

We 
proposed a three-step pipeline to automate the factorization of arbitrary expressions.

\vspace{3pt}
\noindent
\textbf{Step 1. Identify Factorizable Expressions:} Given an expression applied to transform or filter the output of a multi-way join, to automatically analyze whether the expression is factorizable or not, we developed an algorithm to analyze our nested IR, formalized in Alg.~\ref{alg:expression-analysis}. The main idea is to identify subgraphs in the expression that each depends only on the input of one of the source tables \textcolor{black}{or intermediate join nodes}, allowing computations represented by each subgraph to be pushed to the output of the corresponding node or its ancestors, independent of the rest of the expression. Only a factorizable node could be included in more than one subgraph (line 7-13). The algorithm relies on the mapping from input attributes to a node in $e$ to nodes in the join tree (e.g., line 3). This mapping could be obtained by dataflow analysis with depth-first traversal from the node in the expression $e$ to a node in the join graph. 

\vspace{-4pt}
\begin{algorithm}[h]
\small
    \SetKwInOut{getSources}{Objective}
    \getSources{Partition an expression $e$ into independent subexpressions $e_1,...,e_{2m-1}$ with $e_i$ solely relying on node $N_i$ in the join query tree\\}
    \ResetInOut{Output}
    \SetKwInOut{Input}{Input}
    \SetKwInOut{Output}{Output}
    \Input{$e=G(V, E)$: expression $e$ is represented as a computational graph, where $V$ is a set of nodes, i.e., functions or expression oprators in $e$, and $E$ is a set of edges, i.e., dataflows between two nodes; A join query tree $T$ with $2m-1$ nodes: $N_1, ..., N_{2m-1}$; $V_S \subset V$: a set of all source nodes from $e$, with each $v_s \in V_S$ consuming a subset of features output from $T$}
    \Output{$\mathcal{M}$: a mapping from each non-root node $N_1,...,N_{2m-1}$ to a subgraph of $e$ that solely relies on the output of $N_i$ }
    \ForEach{$v_s \in V_S$ that has not been traversed}{      
         $curNode$ $\gets$ \ $v_s$\;
         $N_i$ $\gets$ \ the join node that $v_s$ solely relies on\;
         $V_{i}$ $\gets$ \ $\phi$; $E_{i}$ $\gets$ \ $\phi$; $\mathcal{M}[i]$ $\gets$ \ $\phi$\;
         $C$ $\gets$ consumer nodes $v$ of $curNode$\;
         \ForEach{$v \in C$ that has not been traversed}{%
                  \If{$v$ has only one child node or $v$ is factorizable}{
                  \textcolor{black}{$V_{i}$ $\gets$ \ $V_{i} \cup \{v\}$\;}
                  \textcolor{black}{$E_{i}$ $\gets$ \ $E_{i} \cup \{edge(curNode, v)\}$\;}
                 
                 \If{$v$ has only one child node}{
                    $C$ $\gets$ $C \cup consumers(v)$\;
                 }
                 \Else{\tcp{$v$ is a factorizable node.}  
                    $V_S$ $\gets$ $V_S \cup consumers(v)$\;
                    
                 }
                 }
                 \Else{
                  \tcp{$v$ has more than one child node}
                   $V_S$ $\gets$ $V_S \cup \{v\}$\;
                 }
                 break\;
              }
             $\mathcal{M}[i]$ $\gets$ \ $\mathcal{M}[i] \cup G_i(V_i, E_i)$ \;
          
    }
    \textbf{return} $\mathcal{M}$\;
    \caption{\small Identify factorization possibility of an expression}
    \label{alg:expression-analysis}
\end{algorithm}

\vspace{3pt}
\noindent
\textbf{Step 2. Applying Factorization Optimization:} Once a factorizable expression is identified, it will collect the information required for applying our \textit{Greedy} (Alg.~\ref{alg:greedy-optimization}) or \textit{Genetic} (Alg.~\ref{alg:genetic-optimization}) algorithms, e.g., those listed in Tab.~\ref{tab:sobol-sensitivity}. \textcolor{black}{For cardinality estimation, we apply the histogram-based method~\cite{postgres-cardinality} used in the PostgreSQL DBMS.} 

\vspace{3pt}
\noindent
\textbf{Step 3. Apply the Selected Factorization Plan:} The optimal factorization plan, once selected, will be applied by rewriting our nested IR accordingly. At this step, a factorizable node included into more than one subgraphs will be factorized with each split going to one disjoint subgraph, e.g., MatMul3 is factorized into Group3-1 and Group3-2 respectively, in Fig.~\ref{fig:user-defined-workflow}.

\vspace{5pt}
\noindent
\textbf{Implementation:} We implemented the proposed factorization framework on top of Velox~\cite{pedreira2022velox}, which is a high-performance database engine open-sourced from Meta, developed using C/C++.
Before we apply the factorization, we aggressively push down filter predicates or projections and perform rule-based join ordering optimization~\cite{begoli2018apache}.
\textcolor{black}{After factorization optimization, Velox's physical query optimization, such as parallelism and pipelining, are applied to the fully optimized query plan.}
\textcolor{black}{Velox supports data parallelism by splitting an input table into data splits, and launching a thread over each split. For pipelining and batching, Velox splits the query plan into multiple pipelines, where each pipeline is a sequence of operators that ends with a pipeline breaker.
The processing of one pipeline consists of multiple iterations, where each iteration applies a sequence of vectorized operators to a batch of tuples, with the output of the pipeline being materialized by the pipeline breaker.}

\vspace{5pt}
\noindent
\textcolor{black}{
\textbf{Supported Models:}
Our proposed optimizations for inference workflows can be applied to any pretrained model, as long as the model contains factorizable computations that consume the join outputs. To use our \InferF end-to-end system for automatically identifying factorizable computations and applying our algorithms, the pretrained model or inference workflows must first be translated to our proposed intermediate representation (IR). Models trained on external frameworks, such as PyTorch or TensorFlow, can be converted to ONNX IR~\cite{onnx} format, which further can be translated to our factorizable IR. However, advanced users can always utilize our IR as a Domain-Specific Language (DSL) to represent the model.
}

\section{Experimental evaluation}
\label{sec:experimental-eval}

The experimental evaluation focuses on several questions: \textbf{R1.} How effective are our proposed optimization algorithms, compared to baselines? \textbf{R2.} What are the overheads of the proposed optimization algorithms and IR analysis strategy? \textbf{R3.} What are the factors that impact the speedup of inference workflows due to factorization?
We will also summarize the key observations at the end of the section.

\subsection{Setup and Environments}


\subsubsection{Datasets}
\label{sec:datasets}
Our experimental study involved \textcolor{black}{six} datasets, including:

\noindent
\textbf{The IMDB  dataset}~\cite{job} consists of $21$ tables such as titles, genres, release dates, ratings, cast, and crew. The number of rows in various tables ranges from $4$ to $36{,}244{,}344$. 

\noindent
\textcolor{black}{\textbf{The TPC-DS}~\cite{tpcds-data} is a benchmark dataset to model the decision support system of a retail product. We used this dataset with a scale factor of 1, which contains 24 tables. 
}

\noindent
\textbf{The Epsilon dataset} is a well-known dataset from the Pascal data challenge~\cite{sonnenburg2008pascal}, with $2{,}000$ features and $0.5$ million of tuples. 

\noindent
\textbf{The Bosch dataset}~\cite{mangal2016using} is a well-known manufacturing dataset that represents measurements of parts as they move through Bosch's production lines, having $968$ features and $1.18$ millions of tuples.

\noindent
{\textbf{Expedia Dataset~\cite{projecthamlet2021}} is a hotel ranking prediction dataset with three tables: \textit{listings}, \textit{hotels}, and \textit{searches}. In total, it contains 991{,}102 rows and 28 features. We used this dataset with Scale{=}10.}

\noindent
{\textbf{Flights Dataset~\cite{projecthamlet2021}} is a flights classification dataset which has 4 tables: \textit{routes}, \textit{airlines}, \textit{sairports}, and \textit{dairports}. In total, it contains 73{,}452 rows and 20 features. We used Scale{=}10 for this dataset also.}

\vspace{-5pt}
\subsubsection{Workloads (Queries).}
\label{sec:workloads}

Unfortunately, the existing multi-way join benchmarks~\cite{job} do not involve AI/ML inferences. Therefore, we developed \textcolor{black}{$80$} inference workflows that apply various AI/ML models to features generated by joining $2$ to $20$ tables on the above datasets, organized in two categories.

\vspace{3pt}
\noindent
\textbf{Category 1. Factorization over multi-way joins along foreign key relationships:}
\textcolor{black}{Star schema queries are widely used in data warehouse environments, particularly for decision support systems~\cite{star-schema-1}. Reflecting this realistic query distribution, our workloads in this category primarily consist of star schema queries, with $\approx20\%$ 
of cases introducing additional dimension-dimension joins.}
We constructed (1) $50$ queries with varying numbers of tables synthesized from the Epsilon and Bosch datasets, and varying DNN model architectures and (2) $25$ queries on the IMDB dataset. 

\textbf{(1) A synthetic workload with $50$ queries:} 
We sampled $20$ tables from these two datasets by varying i) the number of columns between $5$ and $500$ and ii) the number of rows between $100$ and $1$ million. Each table has a primary key and may have zero or more foreign keys. Feature column values were selected from the Epsilon and Bosch datasets, while primary key/foreign key values were randomly generated. $50$ queries are designed to join $2$ to $20$ tables \textbf{through the primary key-foreign key relationships}, with the join output passed to a randomly sampled feed-forward neural network (FFNN) model for inferences, with the number of layers randomly sampled from a range of $[1, 5]$ and the number of output neurons in the hidden layers randomly sampled from a range of $[16, 1024]$. For the final layer, the number of neurons was in the range of $[2, 10]$ for classification tasks and $1$ for regression tasks.

\textbf{(2) An IMDB workload with $25$ queries:} $25$ queries were modified from the Join Order Benchmark (JOB)~\cite{job} by joining $2$ to $20$ tables in each query \textbf{along the primary key-foreign key relationship}, with the join output sent to an FFNN model randomly sampled in a way similar to the synthetic workload. 

\noindent
\textcolor{black}{\textbf{Category 2. Factorization of Analytical Queries:} We will evaluate five analytical queries on TPC-DS, IMDB, Expedia, and Flights. The underlying join queries are all along the primary key-foreign key relationships}
(Q1) is a two-tower query on TPC-DS dataset that joins ten tables in its store sales ER-diagram (excluding $promotion$ table). All item-related attributes and customer-related attributes are passed to two separate embedding models. \textcolor{black}{Each model comprises five sequential layer groups, with each group containing a fully connected layer, a batch normalization layer, and a ReLU activation function. A third model on top of both embedding models generates a customer-item recommendation score using cosine similarity.} (Q2) predicts the profit amount \textcolor{black}{using a three-layer deep neural network model} on TPC-DS dataset joining all $14$ tables in its web sales ER-diagram. \textcolor{black}{The regressor model consists of three sequential layers, each comprising a fully connected layer followed by a ReLU activation.}
 (Q3) joins all the tables in the IMDB dataset except the $char\_name$ table and predicts a popularity score for various actors in various movies. \textcolor{black}{Movie and actor features are encoded using a model architecture similar to the feature embedding models in Q1, and the resulting embeddings are input to the deep neural network model as used in Q2.}
 \textcolor{black}{(Q4) joins hotel and search tables with listings table in the \textit{Expedia} dataset, followed by min-max scaling and onehot encoding of numerical and categorical columns, respectively. Later, we apply a logistic regression model to classify the \textit{promotion\_flag} attribute. (Q5) joins airline, source airport, destination airport tables with routes table in the \textit{Flights} dataset and apply min-max scaling and onehot encoding similarly to Q4. Later, a random forest classifier model predicts the \textit{codeshare} attribute for each joined tuple. 
 }

\subsubsection{System Environments}

The following environment was used for experiments: an Ubuntu Linux machine with 48 CPU cores (Intel Xeon Silver 4310 CPU 2.10GHz) and 125GB memory size. 

\subsection{Baselines}
\label{sec:baselines}

Existing publicly available factorized ML systems~\cite{factorize-lmfao, factorize-join, factorize-joinboost, factorize-la, li2019enabling, kumar2015learning} do not support the inference process of most AI/ML models used in this work.
In addition, we did not find any systems that optimize the factorization of general computations over multi-way joins, except a simple and brief discussion in~\cite{factorize-la, kumar2015learning, li2019enabling}. Therefore, our evaluation mainly considers the following baselines.

\begin{itemize}[noitemsep, leftmargin=*]
\item \textbf{Regular Join (No push-down)}: This baseline simply joins multiple data silos based on the equality of the join key, and each joined record includes all features of the dataset.
\item \textbf{Full Factorization}: This approach implemented a default factorization strategy by pushing down $f_i$ to $N_i$ for $i=1,...,2m-1$. 
\item \textcolor{black}{\textbf{Morpheus Extended}: We extended Morpheus~\cite{factorize-la} by applying their heuristics to every join node while traversing the join tree in level order in both bottom-up (Morpheus BU) and top-down (Morpheus TD) manner.}
\textcolor{black}{Their heuristics are based on the \textit{feature ratio} and \textit{tuple ratio} between joined tables to guide factorization decisions,
recommending factorization when \textit{tuple ratio} is $\geq 5$ and \textit{feature ratio} is $\geq 1$.}
\item \textcolor{black}{\textbf{FL Extended}: We also extended FL~\cite{kumar2015learning} with our partial factorization by adapting their cost model for two-way join to examine every join node in both bottom-up (FL BU) and top-down (FL-TD) orders.}
\textcolor{black}{This approach models query cost as a combination of I/O cost and CPU cost, based on block accesses during joins and the number of input tuples and features.}
\item \textbf{Genetic (Ours)}: An implementation of Alg.~\ref{alg:genetic-optimization}. \textcolor{black}{To tune the hyper-parameters, we picked five inference queries with \textit{$3$-way}, \textit{$6$-way}, \textit{$10$-way}, \textit{$15$-way}, and \textit{$20$-way} joins and applied a random search over a search space that included the following values of hyper-parameters: \textit{generation count} — $\{10, 15, 20, 25, 30, 40, 50, 75, 100\}$; \textit{population size} — $\{50, 100, 150, 200, 250, 300, 400, 500\}$; and the number of \textit{new chromosomes per generation} — randomly selected between $15\%$ and $25\%$ of the \textit{population size}. Balancing the accuracy of the generated plans with the optimization latency, we set the values of the hyperparameters as follows in all experiments: \textit{generation count} = $20$, \textit{population size} = $150$, and \textit{new chromosomes per generation} = $25$.} 
\item \textbf{Greedy (Ours)}: An implementation of Alg.~\ref{alg:greedy-optimization}.

\end{itemize}

\noindent
\textcolor{black}{All of the above baselines were implemented in the open-source execution engine Velox~\cite{pedreira2022velox}. We also implemented the complex analytical queries described in category 2 of Section \ref{sec:workloads} in four different in-database ML systems, including EvaDB (v0.3.10)~\cite{evadb, kakkar2023eva}, MADLib (v2.1.0)~\cite{madlib}, PySpark (v3.5.4)~\cite{pyspark} through UDF, and DL-Centric system, where data is stored in the Postgres database (v14.15), and query output is loaded into PyTorch (v2.7.1) using ConnectorX (v0.4.3)~\cite{connectorx}.
We compared the performance of these systems to baselines implemented in Velox.}

\subsection{R1. Optimization Effectiveness}
\label{sec:evaluation-optimizers-ir}

In this section, we evaluated the effectiveness of our proposed optimizers by comparing the execution latency of different baselines.
\textcolor{black}{All query latency values reported for each baseline in this section include both the optimization time and the query execution time.}

\subsubsection{Category 1} We start from evaluating the optimization of factorization over multi-way equi-joins along the foreign key relationships using three workloads described in Sec.~\ref{sec:workloads}.

\begin{figure}[ht]
\centering
\includegraphics[width=0.48\textwidth]{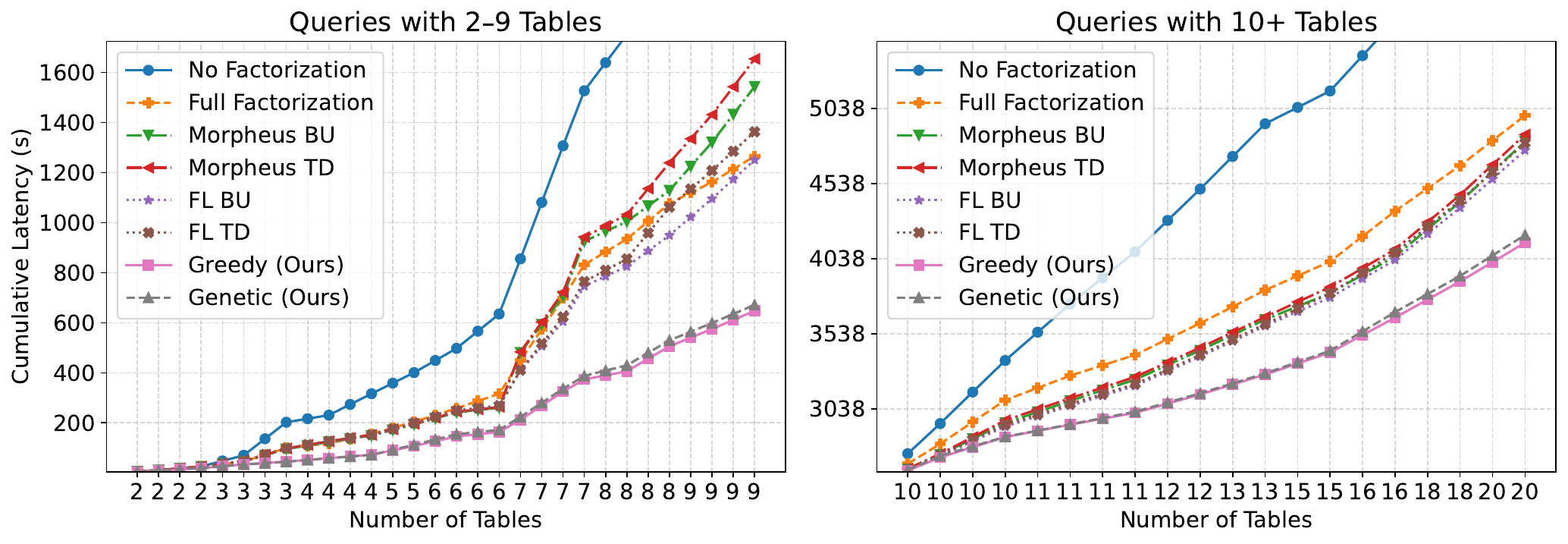}
\caption{\small {Cumulative End-to-End Latency Comparison on $50$ Queries over  Datasets Synthesized from Bosch and Epsilon.} 
}
\label{fig:optimizer-synthetic}
\end{figure}

\noindent
\textbf{(1) Synthetic Workload:}
Fig.~\ref{fig:optimizer-synthetic} depicts the execution latency of the proposed optimizers and the baselines on $50$ queries. Queries are sorted in ascending order of the number of tables and the latency of the no-factorization baseline. Results indicate that our \textit{Greedy} and \textit{Genetic} algorithms have similar performance, and they achieved up to $12.2\times$, $4.9\times$, $4.6\times$, $4.7\times$, $4.6\times$, and $4.7\times$ speedup against \textit{No Factorization}, \textit{Full Factorization}, \textit{Morpheus BU}, \textit{Morpheus TD}, \textit{FL BU}, and \textit{FL TD} baselines, respectively. Furthermore, the increasing speedup achieved by the proposed optimizers as query complexity grows — particularly in queries with many joins - underscores the importance of partial factorization in multi-way join queries. Between the two proposed optimizers, \textit{Genetic} Optimizer introduces performance drop in a few queries due to the inherent randomness of a genetic algorithm, while our proposed \textit{Greedy} Optimizer consistently incurs the optimal execution latency.

\begin{figure}[ht]
\centering
\includegraphics[width=0.48\textwidth]{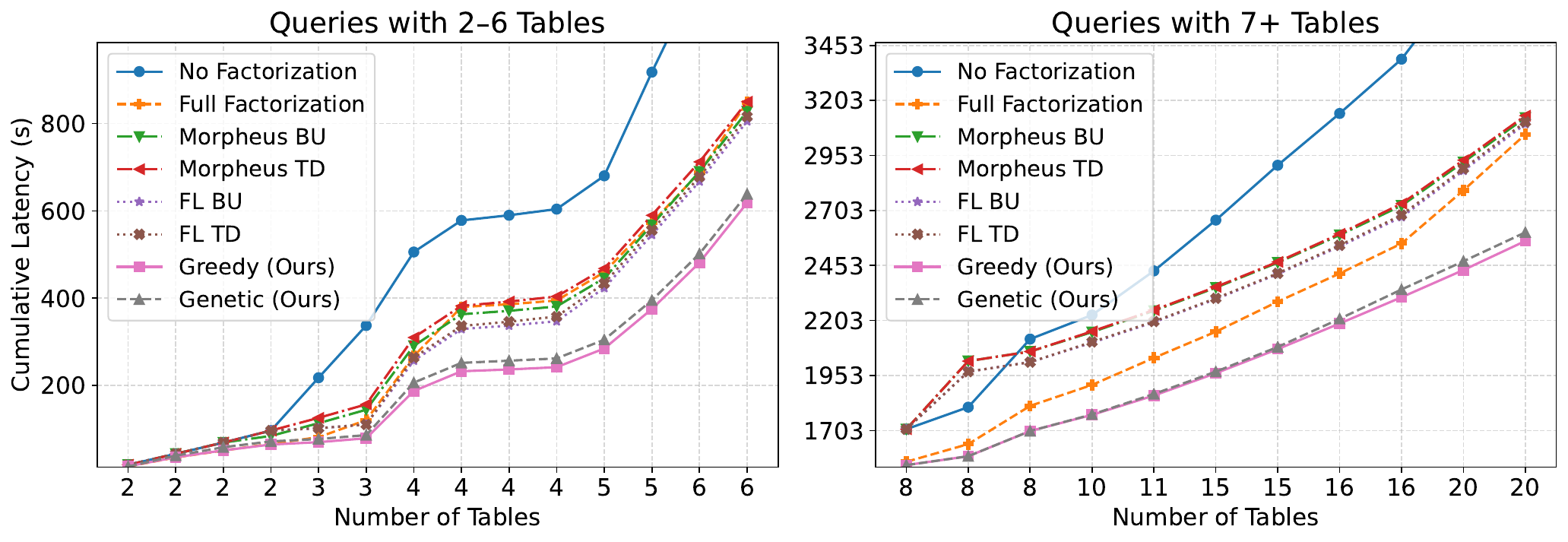}
\caption{\small {Cumulative End-to-End Latency Comparison of Factorization Optimizers on $25$ Queries over IMDB Dataset}}
\label{fig:optimizer-imdb}
\vspace{4pt}
\end{figure}

\noindent
\textbf{(2) IMDB workload:} 
Queries in Fig.~\ref{fig:optimizer-imdb} are sorted similarly to the synthetic queries in Fig.~\ref{fig:optimizer-synthetic}.
The results showed that \textcolor{black}{both \textit{Greedy} and \textit{Genetic} optimizers achieve similar execution speedup, outperforming \textit{No Factorization}, \textit{Full Factorization}, \textit{Morpheus Extended}, and \textit{FL Extended} baselines by up to $22.5\times$ and $4.4\times$, $7.7\times$, and $6.5\times$, respectively.}
Top-down and bottom-up traversals of the \textit{Morpheus Extended} and \textit{FL Extended} baselines display similar performance, with only marginal differences observed in a few high-join queries. Therefore, in the following sections, we use ``Morpheus'' and ``FL'' to refer to the variants with the best performance. 

\vspace{-4pt}
\subsubsection{Category 2.}
\label{sec:cmp-indb-ml}

\begin{figure}[ht]
\centering
\vspace{-8pt}
\includegraphics[width=0.47\textwidth]{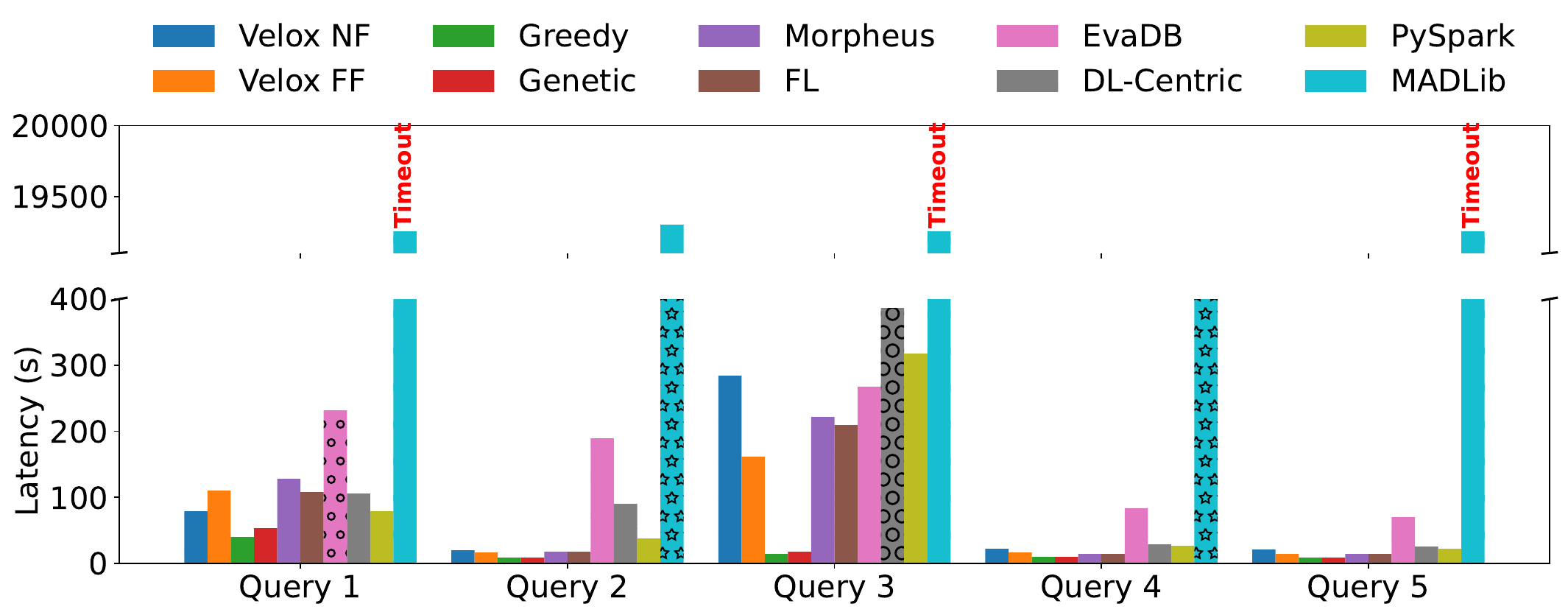}
\vspace{-10pt}
\caption{\small \textcolor{black}{Query Latency Comparison of InDB ML Systems}}
\label{fig:eval-indb-ml}
\vspace{-4pt}
\end{figure}

\noindent
\textbf{Comparison with In-DB ML Systems}
In this section, we compare the performance of the proposed system to \textcolor{black}{various In-DB ML system baselines described in Sec.~\ref{sec:baselines}}.
\textcolor{black}{Fig. \ref{fig:eval-indb-ml} illustrates the performance comparison of all baselines for five complex analytical queries. In Q1, Q2, Q3, Q4, and Q5, \textit{Greedy} optimizer outperforms the best of No Factorization, Full Factorization, and Morpheus/FL, within Velox by $2.0\times$, $1.95\times$, $11.3\times$, $1.7\times$, and $1.6\times$, respectively. For these queries, the speedups achieved by our \textit{Greedy} optimizer compared to the best of other in-DB ML systems are $2.0\times$, $4.7\times$, $18.7\times$, $2.8\times$, and $2.5\times$, respectively. \textit{MADLib} took \textit{$19,216$ seconds} and \textit{$1,446$ seconds} to complete Q2 and Q4 respectively, while  Q1, Q3 and Q5 did not finish execution within $10$ hours.
}
\textcolor{black}{MADLib is slow in linear algebra operations due to its higher per-tuple RDBMS processing overheads and the I/O time for materializing output tables~\cite{cmp-sys-la}.}
The \textit{Genetic} optimizer achieves near-identical performance to the \textit{Greedy} optimizer across all queries, exhibiting only marginally worse latencies in Q2 and Q3. \textcolor{black}{\textit{Morpheus} and \textit{FL} baselines perform comparably to \textit{Full Factorization} in Q1, Q2, Q4, and Q5, but choose inefficient plans for Q3, leading to a substantial increase in latency.} \textit{DL-Centric} approach suffers from the additional data transfer latency in Q2 and Q3 as these queries generate large output data for inference. \textit{EvaDB} exhibits inconsistent performance across queries, with higher latencies observed in those involving more complex models.

\vspace{-4pt}
\subsection{R2. Optimization Overhead}
\label{sec:overhead-analysis}
In this section, we evaluate the runtime overhead of factorization optimization algorithms, factorization analysis, and offline parameter tuning and cost factor analysis.

\begin{figure}[ht]
\centering
\includegraphics[width=0.47\textwidth]{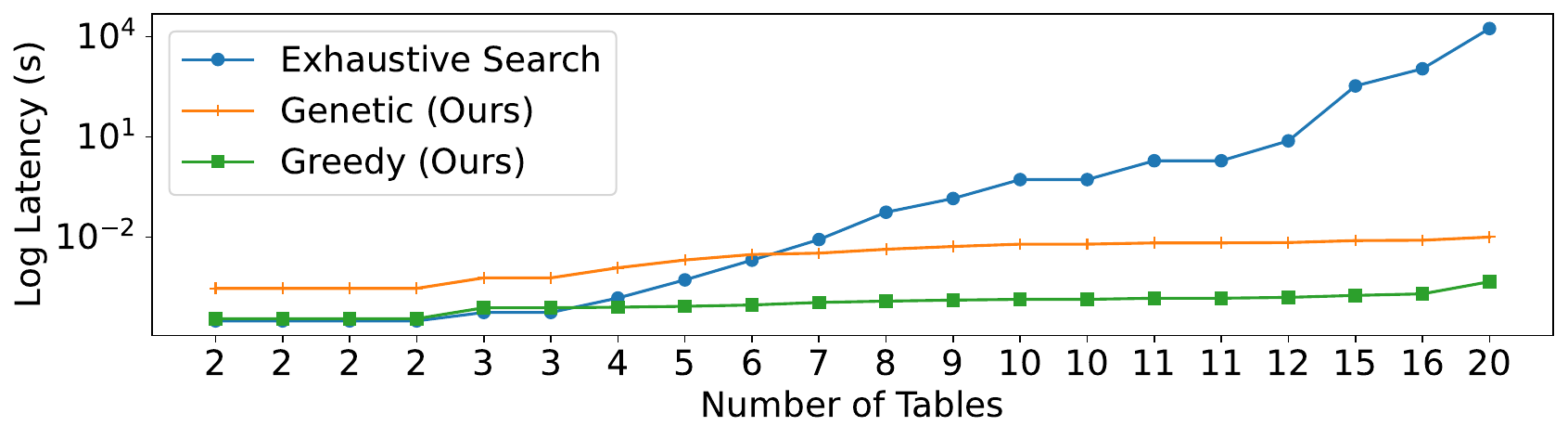}
\vspace{-12pt}
\caption{\small Optimizer Overheads Comparison on IMDB Queries}
\label{fig:optimizer-overhead}
\vspace{-6pt}
\end{figure}

\noindent
\textbf{Overhead of Proposed Optimizers:}
We quantify the additional runtime cost introduced by our \textit{Greedy} and \textit{Genetic} optimizers, comparing them against an exhaustive search strategy that enumerates all possible factorization plans and evaluates their performance.
Fig. \ref{fig:optimizer-overhead} depicts the runtime overhead of the optimizer techniques for the $25$ queries in the IMDB workload (Sec. \ref{sec:evaluation-optimizers-ir}). We observe that the runtime overhead of \textit{Exhaustive Search} increases exponentially with the number of tables, exhibiting an explosive growth once the number of tables exceeds $8$. In contrast, \textit{Greedy} optimizer exhibits significantly lower and practically negligible computational overhead, even beyond $8$ tables.
The \textit{Greedy} optimizer
avoids exploring the search space of possible plans, instead optimizes by iterating over the nodes in the join tree using a node-level cost function. The number of nodes in the join tree is significantly smaller than the total number of possible factorization plans, making the \textit{Greedy} optimizer significantly faster. The runtime of the \textit{Genetic} optimizer is \textcolor{black}{$8.1\times$ to $21.7\times$ higher} than that of the \textit{Greedy} optimizer, but it stabilizes beyond a certain threshold. This stability threshold depends on input
hyperparameters defined in Section \ref{sec:baselines}
which bound the number of chromosomes explored and the number of crossover and mutation steps in a very large.

\noindent
\textbf{Overhead of Factorization Analysis:}
\textcolor{black}{
We found that the factorization analysis based on Alg.~\ref{alg:expression-analysis} introduces negligible overhead of less than $1$ millisecond for most of the queries. 
}

\noindent
\textbf{Offline One-Time Overhead Analysis:}
\label{sec:eval-offline-overhead}
\textcolor{black}{While parameter tuning for the \textit{Genetic} optimizer using random search as described in Sec.~\ref{sec:baselines} takes \textcolor{black}{$864$ seconds}, the sensitivity analysis of cost factors in our \textit{Greedy} optimizer as described in Tab.~\ref{tab:sobol-sensitivity}, 
implemented using the SALib library~\cite{salib} 
only takes  \textit{1.5 seconds}.}
Training the \textit{linear} and \textit{logistic regression} models for the \textit{Genetic} and \textit{Greedy} cost models takes \textit{130.4 seconds} and \textit{93.8 milliseconds}, respectively.

\vspace{-2pt}
\subsection{R3. Factors Impacting Factorization Speedup}
\label{sec:factors-analysis}
In this section, we analyze the impact of various factors, including \textbf{model complexity}, \textbf{join complexity}, \textbf{aggressive aggregation}, \textbf{join order},  on the speedup achieved by the factorization optimization strategy. We also investigate how cost factors affect the accuracy of the \textit{Greedy} algorithm's node-level cost function.



\begin{figure}[ht]
\centering
\includegraphics[width=0.48\textwidth]{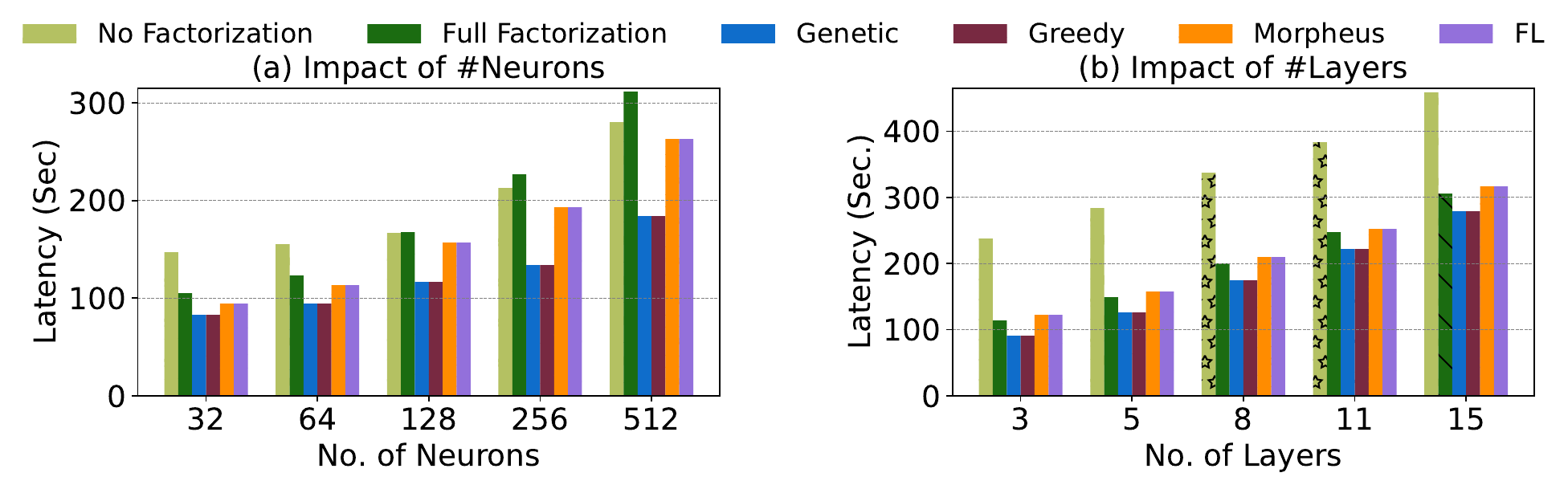}
\vspace{-16pt}
\caption{\small Impact of Model Complexity on Factorization}
\label{fig:eval-as-neuron-layer}
\vspace{-10pt}
\end{figure}

\begin{figure}[ht]
\centering
\includegraphics[width=0.48\textwidth]{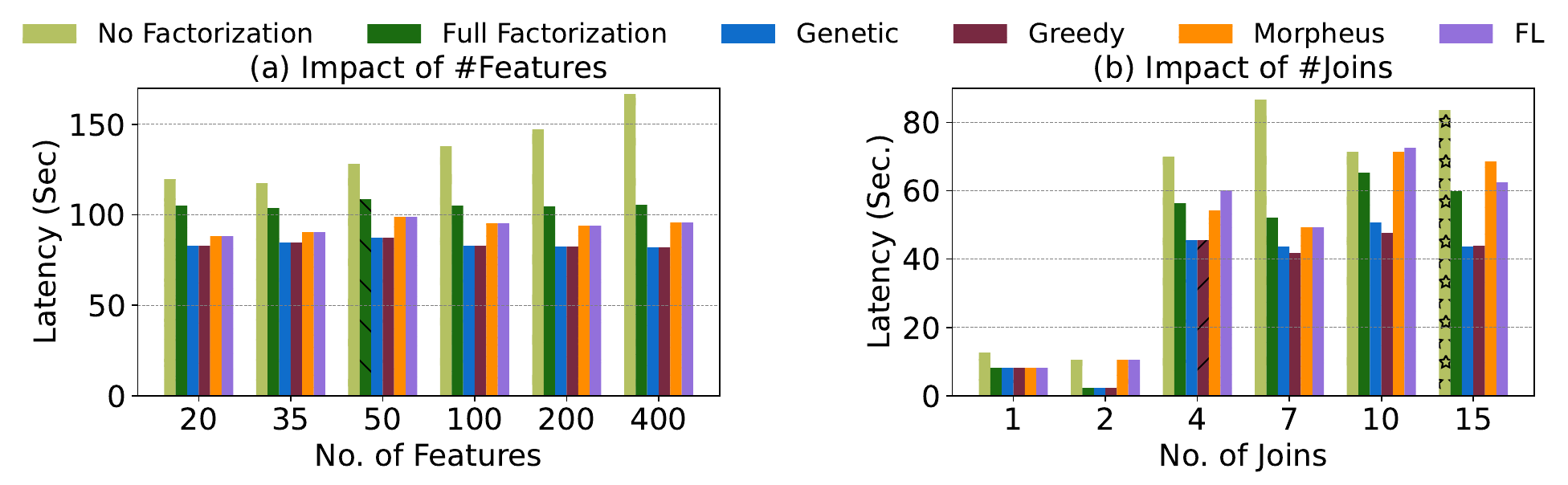}
\vspace{-14pt}
\caption{\small Impact of Query Complexity on Factorization}
\label{fig:eval-as-features-join}
\vspace{-14pt}
\end{figure}

\begin{figure}[ht]
\centering
\includegraphics[width=0.48\textwidth]{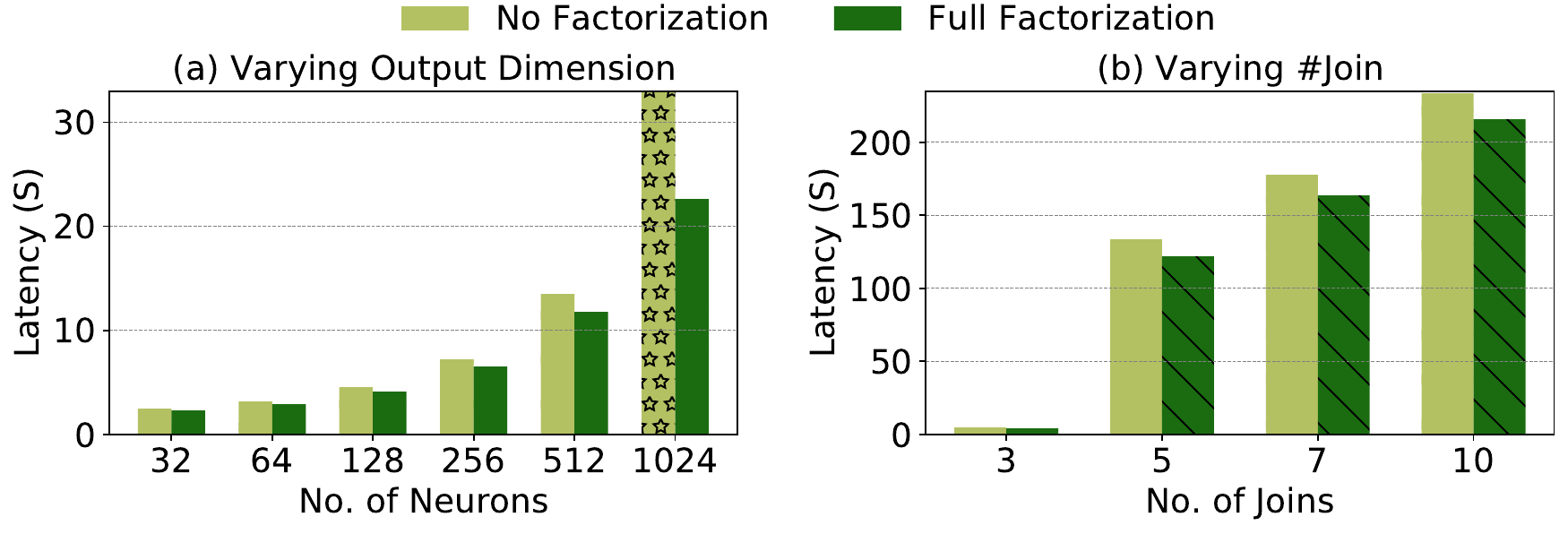}
\vspace{-16pt}
\caption{\small Impact of Aggregation Push Down on Speedup}
\label{fig:eval-impact-aggr-push}
\end{figure}

\noindent
\textbf{Impact of Model Complexity and Join Complexity:}
Fig.~\ref{fig:eval-as-neuron-layer} illustrates how model complexity (e.g., the number of neurons of the factorized hidden layer and the number of hidden layers of FFNN models) influences the speedup achieved through factorization. 
As shown in Fig.~\ref{fig:eval-as-neuron-layer}(a), increasing the number of neurons increases the tuple size after factorization and reduces the I/O cost saving, shrinking the speedup brought by factorization. As the number of neurons increases, the performance of the \textit{Full Factorization}, \textit{Morpheus}, and \textit{FL} baselines degrades significantly, often approaching or even falling below that of the \textit{No Factorization} baseline. However, our \textit{Greedy} and \textit{Genetic} optimizers consistently maintain the speedup by selecting plans that avoid size increase in intermediate results. Fig.~\ref{fig:eval-as-neuron-layer}(b) illustrates the impact of the number of hidden layers on factorization speedup. As the number of hidden layers increases, although the dominance of non-factorized computations over the end-to-end latency affects the relative speedup against \textit{No Factorization}, our \textit{Greedy} and \textit{Genetic} optimizers consistently maintain the relative speedup against other baselines.

Fig.~\ref{fig:eval-as-features-join} illustrates how join complexity (i.e., number of tables to be joined and the dimension of feature vectors input to the AI/ML model) influences the speedup achieved through factorization.
We varied the size of the feature vectors, which are output from a join query over the \textcolor{black}{synthetic dataset} and input to an FFNN model \textcolor{black}{composed of three FC layers with 32, 16, and 2 output neurons, respectively}. As shown in Fig. \ref{fig:eval-as-features-join}(a), speedup achieved through factorization increases as feature size increases because factorization will achieve more I/O cost saving given that the factorized feature size is determined by the FFNN model, which is fixed. 
We further vary the total joins from $1$ to $15$ in a query over \textcolor{black}{IMDB dataset} with a fixed FFNN model \textcolor{black}{composed of three fully connected (FC) layers with 32, 16, and 2 output neurons, respectively}. As shown in Fig.~\ref{fig:eval-as-features-join}(b), incrementing the number of joins will increase the ratio of joining latency to the end-to-end latency and thus reduce the ratio of inference computation to the end-to-end latency. This limits (or shrinks) the performance gain of factorization, which is mainly achieved by avoiding redundant inference computation. 

\begin{table}[H]
\vspace{-8pt}
\caption{\small Latency Comparison for Different Join Orders in Seconds}
\vspace{-12pt}
\label{tab:eval-join-order-decouple}
\scriptsize
{
\begin{center}
    \begin{tabular}{|p{0.6cm}|p{2.1cm}|p{2.4cm}|p{1.8cm}|}
\hline
Query & Optimal Order W/O Factorization & Optimal Order W/ Optimal factorization & Best of Remaining Plans \\ \hline
1 & 6.60 & \textbf{4.11} & 5.16     \\ \hline
2 & 13.58 & \textbf{6.83} & 6.90     \\ \hline
\end{tabular}
\end{center}
}
\vspace{-10pt}
\end{table}

\noindent
\textbf{Impact of Aggregation Push Down:}
We compared two aggregation strategies: \textbf{(1) W/ Aggr Push}, which aggregates output from factorized computations immediately after they are pushed down to a join node, as discussed in Sec.~\ref{sec:factorize-optimize-prob}; \textbf{(2) W/O Aggr Push}, where all factorized model output of all data silos are lazily aggregated at the root node, on \textit{4}-way to \textit{11}-way join queries over \textcolor{black}{IMDB} dataset using FFNN models, with varying model complexity (on a \textit{5}-way join query) and join complexity. As illustrated in Fig.~\ref{fig:eval-impact-aggr-push}), \textit{W/ Aggr Push} achieved \textcolor{black}{up to $1.45\times$ speedup} compared to \textit{W/O Aggr. Push}, since it reduces the tuple size as discussed in Sec.~\ref{sec:factorize-optimize-prob}. \textit{W/ Aggr Push} is beneficial irrespective of model and join complexity. 
It is adopted in our implementation for all other experiments. 

\noindent
\textbf{Impact of Join Order Decoupling:}
As discussed in Sec. \ref{sec:decoupled-join-factorize}, we decoupled the join order optimization from the optimization of factorization plans. We compared the effectiveness of this decoupling strategy to a strategy that exhaustively enumerates and evaluates the performance of all combinations of join orders and factorization plans, using one \textit{3}-way join query and one \textit{4}-way join query \textcolor{black}{over Epsilon dataset using an FFNN model composed of three FC layers with 32, 16, and 2 output neurons, respectively}
As shown in Tab. \ref{tab:eval-join-order-decouple}, we found that optimal factorization plans of the optimal join order outperformed other baselines. For both queries, different join orders having similar execution latency before factorization also lead to similar latency after separate factorization optimization. The experimental analysis concludes that decoupling the join order optimization problem is a reasonable strategy.

\vspace{3pt}
\noindent
\textcolor{black}{\textbf{Impact of Cost Factors on Greedy Cost Function:}}
To evaluate the impact of the cost factors identified via Sobol's analysis on the accuracy of the \textit{Greedy} optimizer's cost model, we split the dataset of 500 samples used to derive the cost model into training ($80\%$) and test ($20\%$) sets, dropping one factor at a time during training. We found removing each of the top factors in Tab.~\ref{tab:sobol-sensitivity}, such as \textit{cardinality ratio}, \textit{factorized computation cost}, \textit{tuple dimension ratio}, and \textit{depth ratio}, reduces test accuracy from $94.4\%$ to $75.2\%$, $81.3\%$, $83.7\%$, and $90.5\%$, respectively, highlighting their relative importance.

\subsection{Summary of Observations}
\label{sec:observations}
Based on the empirical evaluation of various components of \InferF on various workloads, our observed insights can be summarized as follows. (1) 
Our \textit{Greedy} and \textit{Genetic} strategy outperforms \textit{No Factorization}, \textit{Full Factorization}, \textit{Morpheus Extended}, and \textit{FL Extended} baselines by up to $22.5\times$, $11.3\times$, $15.6\times$, and $14.7\times$, respectively. 
(2) \textcolor{black}{Our proposed \textit{Greedy} and \textit{Genetic} algorithms can greatly benefit execution latency of join queries with multiple fact tables of diverse sizes, which are prevalent in data warehouses and analytical workloads.
} (3) \textcolor{black}{\textit{Greedy} and \textit{Genetic} optimizers achieve similar performance in most scenarios. However, the \textit{Genetic} method incurs higher offline parameter tuning and online optimization overheads. }
(4) \textcolor{black}{A higher number of neurons in the factorized layer of the model slightly reduces the factorization speedup due to an increase in the intermediate join result size. An increased number of joins in an inference query increases the ratio of joining latency to end-to-end latency, shrinking the performance gain of factorization. As the number of features increases in the query, the performance gain of factorization also increases.}
\textcolor{black}{(5) \textit{No Factorization} yields good performance when the output cardinality and tuple size at the root node are smaller than that of other nodes in the join tree.
In contrast, \textit{Full Factorization} becomes beneficial only when the output cardinalities and tuple size at the table scan nodes are smaller than join nodes across the join tree.
However, such strict conditions are rarely satisfied across diverse queries. As a result, these strategies often fall short, highlighting the need for partial factorization guided by our proposed \textit{Greedy} and \textit{Genetic} optimizers.
In addition, cost models used by \textit{Morpheus Extended} and \textit{FL Extended} baselines often fail to make optimal decisions for intermediate joins in multi-way join queries due to their simple assumptions about the size of fact and dimension tables.} 
(6) Early aggregation of factorized computations in a join node speeds up the query execution by \textcolor{black}{$1.45 \times$} by reducing the tuple size of intermediate join nodes.
(7) \textcolor{black}{Our strategy for decoupling join order optimization from factorization optimization is reasonable since the optimal factorization plan of the optimal join order performs the best among all combinations of join orders and factorization plans.} 
\vspace{-2pt}

\section{Related Works}
\label{sec:related-works}

{\textbf{Relationships to existing factorized ML works.} Table~\ref{tab:summary-related-work} summarizes the key differences between \InferF and prior work on factorized ML. A major line of research on factorized ML~\cite{factorize-lmfao, factorize-join, factorize-joinboost, factorize-database, factorize-ac-dc, factorize-faq-ai, factorize-f} focuses on ML training for specific models, such as linear regression, gradient boost tree models, and the learning processes relying on batch gradient descent (BGD). LMFAO~\cite{factorize-lmfao}, AC/DC~\cite{factorize-ac-dc}, and F~\cite{factorize-f, factorize-join} abstract the feature interaction of the training process as a collection of aggregation queries over a join graph. While effective in training contexts, this formulation does not naturally extend to inference workloads, which often lack an inherent batch aggregation structure. FL~\cite{kumar2015learning} proposed cost models specific to BGD-based learning process to make coarse-grained decisions such as whether to materialize, stream, or factorize.
Other systems, such as Morpheus~\cite{factorize-la} and MorpheusFI~\cite{li2019enabling}, focuses on linear algebra (LA) with and without feature intractions over normalized data,  while SDQL~\cite{shaikhha2022functional} focuses on unified relational algebra (RA) and LA grounded in semiring semantics. They leverage factorization-based rewrite rules to provide optimization opportunities for workloads involving LA. However, these approaches do not address the challenges of exploring the vast space of semantically equivalent rewrites, such as deciding which factorized sub-computations to push down to specific nodes within a join tree, nor do they support arbitrary inference workflows that cannot be expressed solely in terms of LA operators or semiring abstractions.} 

\textcolor{black}{Since our work focuses on the inference process, the training-specific optimizations, such as semi-ring and batch-aggregation representation, and assigning aggregation queries to views based on count of group by attributes in each view, as proposed/adopted in existing works such as F~\cite{factorize-join}, LMFAO~\cite{factorize-lmfao}, and JoinBoost~\cite{factorize-joinboost}, are orthogonal with our work, and not supported in our system. 
}

\begin{table}[t]
\caption{\small {Comparison with existing Factorized ML systems}}
\label{tab:summary-related-work}
\scriptsize
{
\begin{center}
\color{black}
    \begin{tabular}{|c|p{1cm}|p{1.3cm}|p{1.7cm}|p{1.2cm}|}
\hline
 & Target Workloads & Fine-grained Push-Down &Cost Model & Arbitrary Workflow \\\hline
 LMFAO~\cite{factorize-lmfao} & Training &  No & No & No  \\ \hline
 F\& F/SQL~\cite{factorize-f, factorize-join} & Training &  No & No& No  \\ \hline
 JoinBoost~\cite{factorize-joinboost} & Training &  No & No& No  \\ \hline
 AC/DC~\cite{factorize-ac-dc} & Training &  No & No& No \\ \hline
 FL~\cite{kumar2015learning} & Training & No & Yes, specific to BGD& No \\
 \hline
 Morpheus~\cite{factorize-la} & LA & No & No& No \\ \hline
 MorpheusFI~\cite{li2019enabling} & Training &  No & No& No  \\ \hline
 SDQL~\cite{shaikhha2022functional} & LA\&RA &  No & No& No, semi-ring \\ \hline
\textbf{\InferF (Ours)} & \textbf{Inference} &  \textbf{Yes} & \textbf{Yes}& \textbf{Yes} \\ \hline
\end{tabular}
\end{center}
}
\end{table}

\noindent
\textcolor{black}{\textbf{Relationships to in-DB ML systems.}} In-database ML systems, such as MADLib~\cite{madlib},  Raven~\cite{park2022end}, EvaDB~\cite{evadb}, \textcolor{black}{GaussML~\cite{gaussml}}, Imbridge~\cite{zhang2024imbridge, zhang2025mitigating}, as well as intermediate representations like Weld IR~\cite{palkar2018evaluating}, Tensor Relational Algebra~\cite{DBLP:journals/pvldb/YuanJZTBJ21, DBLP:journals/sigmod/JankovLYCZJG20, guan2023comparison}, and Lara~\cite{kunft2019intermediate}, LingoDB~\cite{jungmair2023declarative, jungmair2022designing}, 
also support the models used in this work. However, they focus on other SQL-ML co-optimization techniques such as tensor-relational transformation, data-driven model pruning, cross-library loop fusion, etc., and do not incorporate the factorization-aware optimization. There is also a line of works on storage optimization for inference queries~\cite{guan2025privacy, DBLP:journals/pvldb/ZhouCDMYZZ22} and ML-in-DB~\cite{zou2020architecture, zou12pangea, zou2021lachesis}, which are orthogonal to \InferF.
These works may benefit from integrating with our work.


\noindent
\textcolor{black}{
\textbf{Relationships to statistical relational learning.} 
We acknowledge that the ML models studied in this paper, though widely used for predictions on relational data~\cite{park2022raven, gaussml, zhang2025mitigating, madlib, postgresml, bisong2019google, evadb}, have two limitations.
First, they do not attempt to exploit dependencies among relational instances, and any dependencies across instances are ignored~\cite{neville2003statistical}, which may make them unsuitable for certain learning tasks that rely on such correlations, such as identifying relationship between entities.
Second, these models are designed to learn over independent-and-identically-distributed  (i.i.d.)\ datasets; however, the i.i.d.\ assumption is often violated in relational settings due to dependencies across tuples~\cite{dhurandhar2013auto,scalable-relational-learning}.}

\textcolor{black}{
To exploit the cross-tuple dependencies (i.e., Limitation 1), there emerged works on relational learning and statistical relational learning (SRL)~\cite{intro-statistical-rel-learning}, including relational learning with language bias~\cite{scalable-relational-learning}, machine learning with inductive logic programming~\cite{QuickFOIL}, probabilistic relational models~\cite{getoor2001learning}, Markov logic network~\cite{richardson2006markov}, relational Bayesian networks~\cite{jaeger2007parameter}, and propositional approaches~\cite{kramer2001propositionalization, roth2001propositionalization}. 
}

\textcolor{black}{
To mitigate i.i.d.\ violations (Limitation~2) for non-relational models in practice, modern ML pipelines typically employ 
\emph{feature extraction} and \emph{data partitioning} techniques that transform multi-table relational data into approximately independent samples suitable for statistical learning algorithms, where relational attributes are joined and aggregated from related tables to produce flat feature vectors for each prediction target~\cite{kumar2016join, jensen2002linkage, lao2015learning, anderson2013brainwash, zhou2023febench}.
Relational learning is also often applied to extract features for non-relational model~\cite{jensen2002linkage, lao2015learning}.}

\textcolor{black}{
On one hand, we acknowledge that non-IID issues should be addressed upstream during training.
On the other hand, mitigation of this concern is orthogonal to our contribution on inference.
A natural direction for future work is to co-design factorized training and inference procedure that explicitly accounts for dependency in relational data, thereby improving the accuracy while retaining the computational benefits at inference.}

\section{Conclusion}
\label{sec:conclusion}
Recently, optimizing the execution latency of inference queries that nest SQL and AI/ML model inferences has received significant attentions~\cite{park2022end, DBLP:journals/pvldb/YuanJZTBJ21, jankov2019declarative, postgresml, evadb, gaussml, amazon-feature-store, lin2023smartlite}, because those queries are widely used in industry and data science, and shortening the latency of those queries brings better user interaction experiences and more timely decision making.
In this work, we identified the significant gaps between the real-world demands for fully optimizing inference queries that nest multi-way join queries with AI/ML inferences and existing factorized ML works focusing on the learning process and linear algebra operators. We propose \InferF to systematically close the gaps by formalizing a new problem on assigning factorization decisions to query nodes, designing new algorithms such as our \textit{Greedy} and \textit{Genetic} optimizers to search for the optimal factorization plan efficiently, and designing a novel representation for arbitrary inference workflows as analyzable expression graphs to enable fine-grained factorization optimization. Our implementation on Velox~\cite{pedreira2022velox} and experimental studies demonstrate that
\InferF significantly speeds up inference on multi-way join queries \textcolor{black}{by utilizing the fine-grained factorization and push-down benefits.}

\noindent 



\bibliographystyle{ACM-Reference-Format}
\bibliography{references}

\newpage
\clearpage
\appendix 
\begin{table*}
\vspace{-10pt}
\caption{ {Comparison with existing Factorized ML systems}}
\vspace{-10pt}
\label{tab:summary-related-work1}

{
\begin{center}
\color{black}
    \begin{tabular}{|c|p{1.5cm}|p{2cm}|p{1.5cm}|p{2.5cm}|p{4cm}|p{1.5cm}|}
\hline
 & Target Workloads & Fine-grained Pushdown Decisions & Cost Model&Arbitrary Workflow& Automatic Optimization & Max \# joins evaluated\\\hline
 LMFAO~\cite{factorize-lmfao} & Training &  No &No& No, only models that are batches of aggregations &  Aggregation-specific, simple heursitics (groupby attr count) to assign aggs to tables&10\\ \hline
 F~\cite{factorize-join, factorize-f, factorize-database} & Training &  No &No & No, only LR models & No &5\\ \hline
 JoinBoost~\cite{factorize-joinboost} & Training &  No &No& No, only tree-based models & Use LMFAO optimizer &10\\ \hline
 AC/DC~\cite{factorize-ac-dc} & Training &  No &No & No, only learning processes using BGD & Aggregation-specific, aggregation-root mapping based on FD &5\\ \hline
 FL~\cite{kumar2015learning} & Training & No &Model-Specific& No, focusing on BGD-based learning & Coarse-grained decision: factorize, materialize, or stream&11\\
 \hline
 Morpheus~\cite{factorize-la} &LA computations & No &No& No, only known linear algebra operators & Coarse-grained decision: factorize or materialize, using simple rule based on input tables' tuple ratio and feature ratio&4\\ \hline
 MorpheusFI~\cite{li2019enabling} & LA w/ feature interaction in training&  No &No & No, only for LA operators with feature interactions  & Coarse-grained decision: factorize or materialize, using simple rules based on input tables' tuple ratio, feature ratio, and sparsity &4 \\ \hline
 SDQL~\cite{shaikhha2022functional} & RA, LA, and hybrid computations (e.g., co-variance) &  No &No& No, only models represented as semi-ring & Applying rewrite rules in fixed order &6\\ \hline
\textbf{\InferF (Ours)} & \textbf{Inference} &  \textbf{Yes} &\textbf{Yes}& \textbf{Yes} & \textbf{Yes}  &\textbf{20}   \\ \hline
\end{tabular}
\end{center}
}
\vspace{-10pt}
\end{table*}

{\color{black}{
\section{Extended Literature Survey}
\label{sec:literature-survey}
\subsection{Relationships to Factorized ML}
We contrast \InferF with prior \emph{factorized ML} systems along six axes that matter for inference over multi-way joins:
(i) Target workoads, e.g., \emph{training}, \emph{inference};
(ii) support for \emph{fine-grained selective push-down} beyond pushing all computations to base tables;
(iii) existence and purpose of a \emph{cost model};
(iv) support for \emph{arbitrary inference workflows} (incl.\ UDFs and non-linear ops) beyond fixed algebraic forms;
(v) \emph{automatic optimization} across a search space of plans (vs.\ rule/heuristic choices);
and (vi) the \emph{scale of join graphs} studied.
Table~\ref{tab:summary-related-work1} summarizes these dimensions.

\vspace{4pt}
\noindent
\textbf{Aggregation-based, training-centric systems.}
\emph{LMFAO}~\cite{factorize-lmfao} optimizes large batches of group-by aggregates over joins during \emph{training}.
It assigns aggregation roots in a join tree via a greedy heuristic that favors relations covering many group-by attributes (and breaks ties by relation size) to maximize view sharing. Such heuristics do not exist at inference time.
This approach assumes each aggregate admits a fixed attribute and root relationship and targets batch aggregation plans, not inference-time, selective push-down of general computations.
\emph{AC/DC}~\cite{factorize-ac-dc}, and \emph{F, FDB, and F/SQL}~\cite{factorize-join,factorize-f,factorize-database} similarly formulates (convex) learning  and linear regression learning over normalized data as collections of aggregates over joins and uses functional-dependency aware planning for gradient descent.

\emph{FL}~\cite{kumar2015learning} analyzes training-time strategies (factorize, materialize, stream, stream-reuse, partition) and presents a cost model that estimates CPU and I/O costs for batch gradient descent (BGD); the model guides end-to-end training strategy and partition choices under the assumption of using hybrid hash algorithms. Its extension for multi-table join include a problem formalization focusing on the selection of the number of table partitions for implementing the hybrid hash algorithm, not for \emph{fine-grained selective push-down} over multi-way joins or non-linear inference.

\vspace{4pt}
\noindent
\textbf{Learning tree-based models over normalized data.}
\emph{JoinBoost}~\cite{factorize-joinboost} rewrites boosting and random-forest training as SQL over normalized data.
Its optimizer adopts LMFAO-style aggregation-root assignments to avoid denormalization, remaining training-focused and aggregation-specific.

\vspace{4pt}
\noindent
\textbf{Linear-algebra (LA) rewrites over normalized data.}
\emph{Morpheus}~\cite{factorize-la} introduces rewrite rules to execute linear algebra (LA) operators directly over normalized data by mapping denormalized matrix ops to per-relation computations; A coarse-grained decision on whether to factorize is guided by simple table/feature ratio rules and primarily evaluated for two-way joins.
\emph{MorpheusFI}~\cite{li2019enabling} extends this to feature interactions with additional sparsity-aware heuristics.
These systems target known LA operators and provide coarse choices (factorize vs.\ materialize), without a general cost model for enumerating \emph{fine-grained selective push-down} plans on multi-way joins or arbitrary inference workflows.

\vspace{4pt}
\noindent
\textbf{Semiring-based unification.}
\emph{SDQL}~\cite{shaikhha2022functional} unifies relational algebra (RA) and LA, via semiring dictionaries and applies a fixed set of rewrite rules.
While expressive, SDQL does not pose the \emph{plan search} for inference push-down as a cost-based optimization problem nor analyze partial factorization opportunities across a join tree.

\vspace{4pt}
\noindent
\textbf{Positioning and takeaways.}
Across aggregation-centric (LMFAO, AC/DC, F, JoinBoost), LA-centric (FL, Morpheus, MorpheusFI), and semiring-centric (SDQL) lines of work,
prior systems either (1) focus on \emph{training} and model-specific algebra, or (2) rely on heuristics / coarse decisions (e.g., root choice; factorize vs.\ materialize) without exploring the exponential space of \emph{fine-grained selective push-down} plans in multi-way joins, or (3) lack a cost model tailored to inference pipelines with non-linear/UDF operators.
By contrast, \textbf{\InferF} targets \emph{inference}, represents arbitrary workflows as analyzable expression graphs, and \emph{formally} defines the \emph{fine-grained selective factorization and push-down} problem over multi-way joins; we capture both computation and I/O effects, support \emph{group push-down} for aggressive aggregation of intermediate inference results at internal join nodes, prove NP-hardness, decouple factorization from join-order optimization, and introduce \emph{two automatic optimizers} (genetic and greedy) to navigate the search space efficiently.

\vspace{6pt}
\noindent
\textbf{Practical implications for inference over joins.}
For real-world batch scoring and UDF-heavy pipelines, the key gaps left open by training-focused systems are:
(1) identifying which sub-expressions are safely factorizable to specific base or internal join nodes;
(2) quantifying benefits of width reduction from \emph{aggregating} multiple pushed-down sub-results at internal nodes (not just pushing to leaves);
and (3) balancing compute vs.\ join I/O under realistic cardinality estimates in \emph{multi-way} joins.
\InferF directly addresses these gaps with a cost objective that combines per-node compute and propagated I/O effects, enabling principled partial push-down beyond prior heuristic or rule-based strategies.

\subsection{Relationships to other AI/ML (inference) systems over relational data.}

Recently,
a trend of supporting SQL queries nested with AI/ML inference functions within database systems has emerged~\cite{gaussml, kakkar2023eva, postgresml, lin2023smartlite, 
bisong2019google, 
shahrokhi2024pytond, armenatzoglou2022amazon, shaikhha2021intermediate}. This approach eliminates the need to transfer data from databases to ML systems. It not only reduces latency in workloads where data transfer becomes a bottleneck~\cite{guan2023comparison} but also alleviates privacy concerns~\cite{annas2003hipaa, regulation2018general} and operational overhead~\cite{amazon-tco}.

UDF-centric systems such as EvaDB~\cite{kakkar2023eva} and PostgresML~\cite{postgresml} allows inference logic to be encapsulated in user-defined functions. MADLib~\cite{madlib} and PySpark~\cite{pyspark_model_infer} provide partial support for UDF pushdown, but are limited in their handling of arbitrary UDFs. For instance, MADLib cannot push down table-level UDFs like \textit{madlib.mlp\_classification}, while PySpark supports the push-down of MLlib functions but not general UDFs. More importantly, these systems lack the ability to analyze or transform UDF internals, making them unsuitable for deeper co-optimization as required by factorized ML.

Other systems like SimSQL~\cite{luo2018scalable,jankov2019declarative}, SystemML~\cite{boehm2016systemml}, and SystemDS~\cite{boehm2019systemds} embed ML workloads in relational algebra atop big data processing platforms like Hadoop~\cite{white2012hadoop} or Spark~\cite{zaharia2010spark}. MASQ~\cite{paganelli2023pushing,del2021transforming} rewrites a subset of inference operators—including one-hot encoding, linear models, and decision trees—into SQL. 
Raven~\cite{park2022end} combines relational, LA, and ML operators in a unified IR to apply ML-informed rule-based optimization, using a classification-based strategy, and a regression strategy to decide which to choose from three options, ML2SQL, ML2DNN, and no optimization. However, all of these systems do not cover factorized ML techniques and are thus orthogonal to our work.

\subsection{(statistical) relational learning.}
Relational data are ubiquitous in enterprise and scientific applications, yet most ML models, including those studied in this paper, assume input data are organized as flat, independent samples. 
We acknowledge that such models have two key limitations.
First, they do not attempt to exploit dependencies among relational instances, and any dependencies across instances are ignored~\cite{neville2003statistical}, which may make them unsuitable for certain learning tasks that rely on such correlations---for example, identifying the supervisory relationship between Ph.D.\ students and professors based on publication and course information.
Second, these ML models are designed to learn over i.i.d.\ datasets; however, the i.i.d.\ assumption is often violated in relational settings due to dependencies across tuples~\cite{dhurandhar2013auto,scalable-relational-learning}.

To exploit the cross-tuple dependencies (i.e., Limitation 1), there emerged a broad class of works on relational learning and statistical relational learning (SRL)~\cite{intro-statistical-rel-learning}, including but not limited to relational learning with language bias~\cite{scalable-relational-learning}, machine learning with inductive logic programming~\cite{QuickFOIL}, probabilistic relational models~\cite{getoor2001learning}, Markov logic network~\cite{richardson2006markov}, relational Bayesian networks~\cite{jaeger2007parameter}, and propositional approaches~\cite{kramer2001propositionalization, roth2001propositionalization}. However, these learning tasks are not the focus of this work and are therefore outside the scope.

To mitigate i.i.d.\ violations (Limitation~2) in practice, modern ML pipelines typically employ 
\emph{feature extraction} and \emph{data partitioning} techniques that transform multi-table relational data into approximately independent samples suitable for statistical learning algorithms, where relational attributes are joined and aggregated from related tables to produce flat feature vectors for each prediction target~\cite{kumar2016join, jensen2002linkage, lao2015learning, anderson2013brainwash, zhou2023febench}.
Relational learning can also be applied to extract features for non-relational model~\cite{jensen2002linkage, lao2015learning}.
%
To further reduce tuple-level dependencies and avoid information leakage, data splitting strategies such as \emph{entity-based}, \emph{temporal}, and \emph{stratified} train/test partitions are used to separate examples that share relational neighbors or overlapping join keys~\cite{de2020experimental}.

Since this paper focuses on the \emph{inference stage}, we assume that the pretrained models have been produced by pipelines where such relational feature extraction and partitioning techniques have already alleviated i.i.d.\ violations.
Our work is therefore orthogonal and complementary: \textbf{\InferF} optimizes \emph{inference-time computation push-down and factorization} for these pretrained models or workflows involving one or more such models, rather than addressing relational dependency learning during model training.

To bound the scope of non-IID effects in evaluating \InferF, our experiments (Section \ref{sec:experimental-eval}) only use workloads where the joins are primary key–foreign key relationships. 
Our proposed factorized inference scheme
losslessly preserves the outputs of a trained model while eliminating redundant computation. It neither introduces nor corrects statistical biases that may arise from non-IID training data, only accelerates inference on a pretrained model over relational joins.
\textcolor{black}{
A natural direction for future work is to co-design factorized training and inference procedure that explicitly account for dependency in relational data, thereby improving the accuracy while retaining the computational benefits at inference.
}

}}

\section{Evaluation on Python Environment}

Besides the Velox environment, we also implemented our proposed factorization strategy in Python and compared the following baselines in terms of execution latency of inference queries.

\subsection{Baselines}
\begin{itemize}[noitemsep, leftmargin=*]
\item \textbf{Regular Join}: The Python implementation of the \textit{Regular Join} baseline discussed in the main paper.
\item \textbf{Index Join}: This is an optimized join strategy for dimension reduction, in which each joined record only includes the IDs of the joined tuples from both data silos and the joining key. Later, the feature vector is lazily created from the tuples indexed by their IDs, during the inference process.
\item \textbf{Greedy (Ours)}: Python implementation of factorization plans according to the \textit{Greedy} algorithm discussed in the main paper.
\end{itemize}

Since the index join baseline can transform higher-dimensional joins into significantly lower-dimensional ones, the evaluation in this section aims to compare the speedup achieved by \textit{Greedy} optimizer against that of the index join. For this evaluation, we used similarity or range join queries on both synthetic datasets (Epsilon and Bosch). All datasets were stored in the PostgreSQL database.

\subsection{Evaluation Results}
\begin{figure}[hbt]
\begin{subfigure}{0.23\textwidth}
\centering
\includegraphics[width=0.95\textwidth]{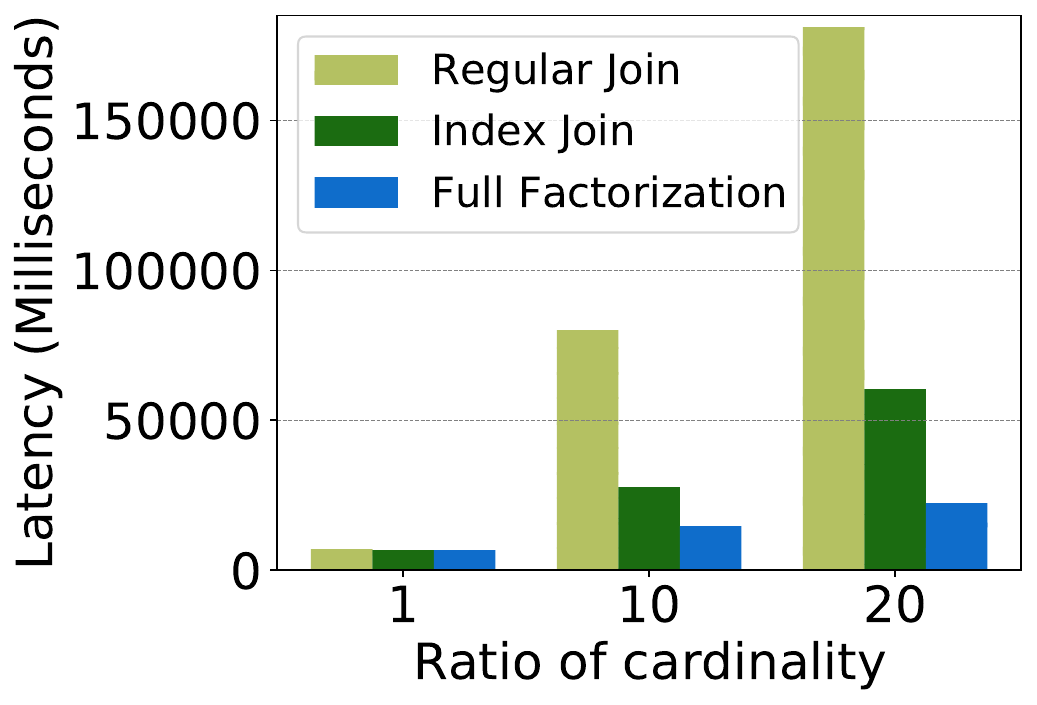}
\caption{2-Partition Dataset}
\label{fig:linear-eps-2-partition}
\end{subfigure}%
\begin{subfigure}{0.23\textwidth}
\centering
\includegraphics[width=0.95\textwidth]{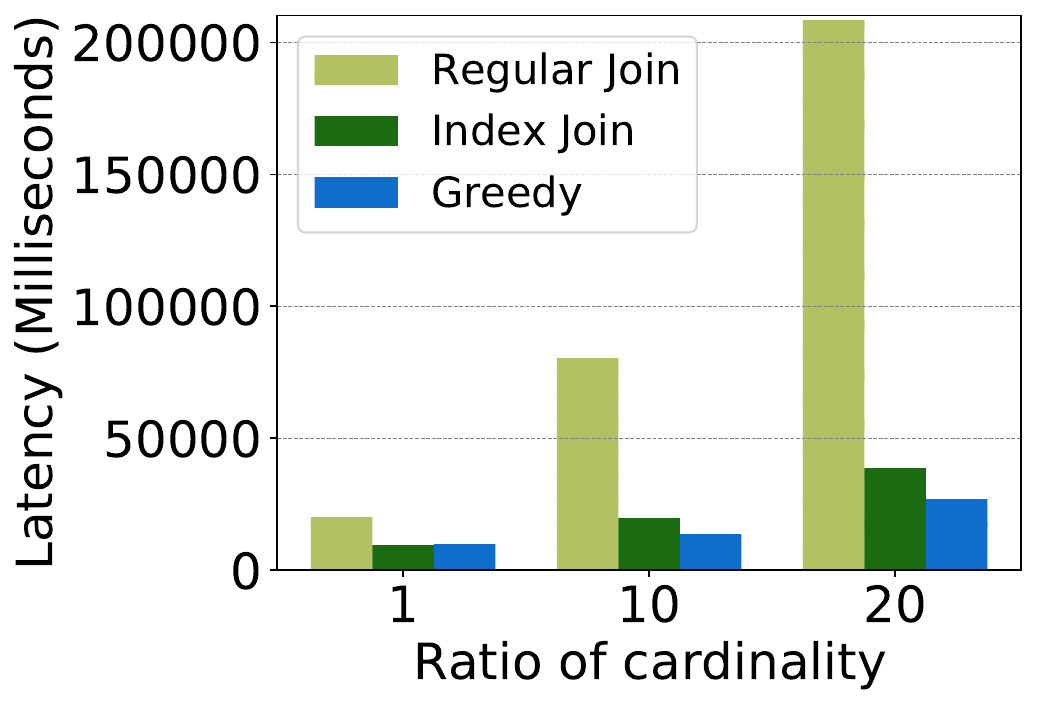}
\caption{5-Partition Dataset}
\label{fig:linear-eps-5-partition}
\end{subfigure}
\begin{subfigure}{0.23\textwidth}
\centering
\includegraphics[width=0.95\textwidth]{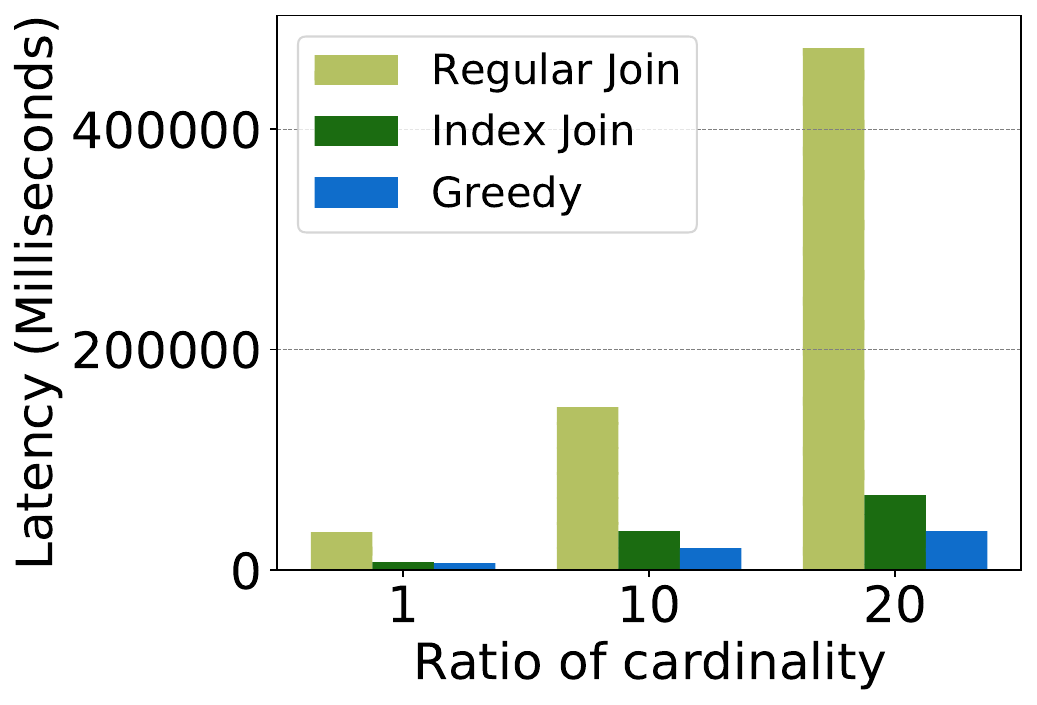}
\caption{10-Partition Dataset}
\label{fig:linear-eps-10-partition}
\end{subfigure}
\begin{subfigure}{0.23\textwidth}
\centering
\includegraphics[width=0.95\textwidth]{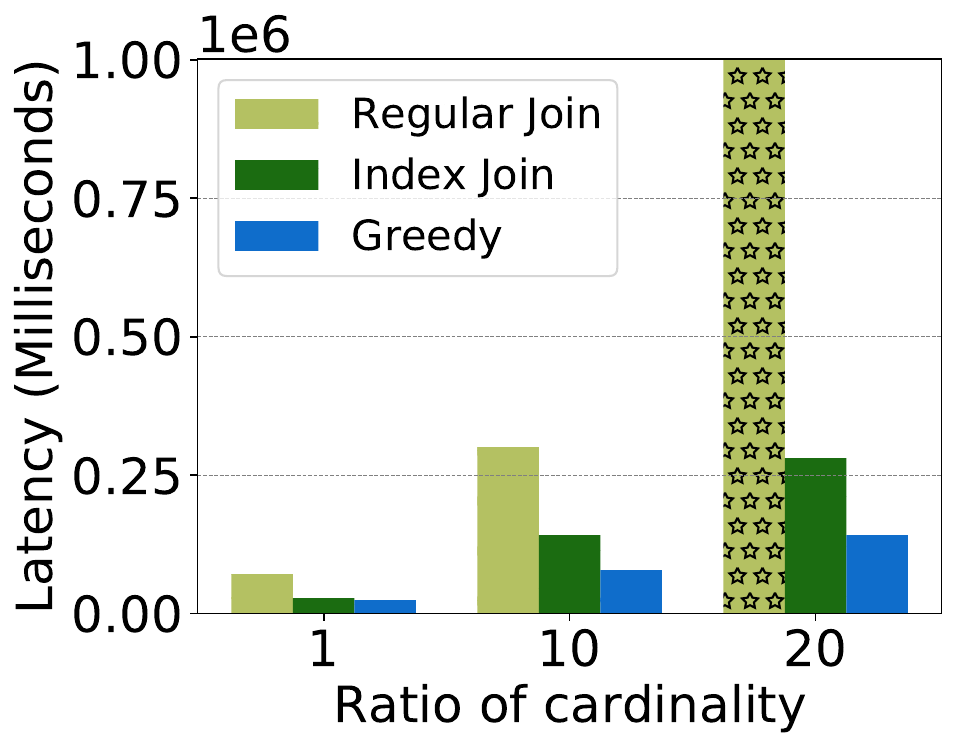}
\caption{20-Partition Dataset}
\label{fig:linear-eps-20-partition}
\end{subfigure}%
\vspace{-5pt}
\caption{
\label{fig:lniear-multi-partition-eps} Evaluating Speedup by Factorization of Neural Networks with Multi-Partition Dataset
}
\vspace{-5pt}
\label{fig:eval-linear}
\end{figure}

\textbf{Evaluating Multilayer Neural Networks}:
Figure \ref{fig:lniear-multi-partition-eps} compares the performance of the baselines on a multilayer neural network, varying both the cardinality ratio (i.e., the ratio of the output cardinality to the input cardinality during join processing) and the number of partitions on the Epsilon dataset. As illustrated in the figure, \textit{Greedy} factorization achieved up to $8\times$ speedup compared to the regular join and exhibits $2.65\times$ speedup compared to the index join. Results on various cardinality ratio indicate that \textit{Greedy} factorization performs significantly better as the cardinality ratio increases (which is the usual case for most join scenarios). Even though the join overhead in the multi-partition datasets is significantly high compared to that in simple two-partition datasets, \textit{Greedy} factorization still can achieve more than $6\times$ speedup compared to regular join and up to $2\times$ speedup compared to index join baseline.
The explanation for the speedup of \textit{Greedy} factorization is that this approach executes the factorized part of the model for $\Sigma_{i=1}^{i=k} |\mathbf{x_i}|$ samples instead of $|z|$ samples where $\Sigma_{i=1}^{i=k} |\mathbf{x_i}| << |z|$. Besides, in the case of \textit{Full Factorization}, the multiplication operation in the decomposed layer is relatively smaller, resulting in a better speedup gain.

\begin{figure}[ht]
\begin{subfigure}{0.23\textwidth}
\centering
\includegraphics[width=0.98\textwidth]{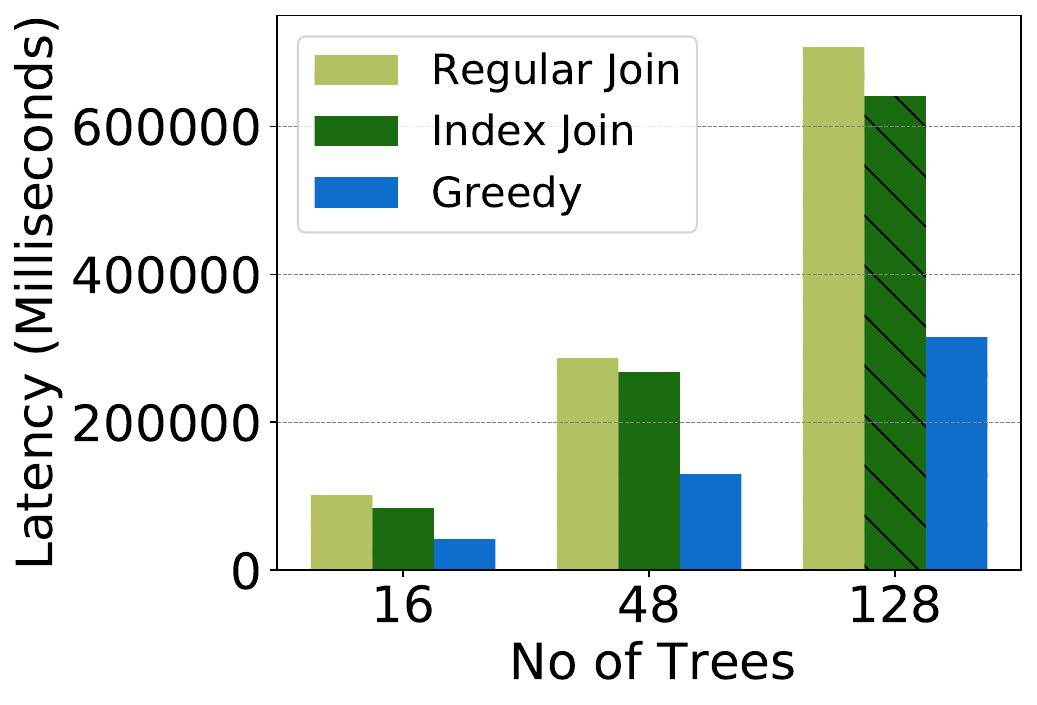}
\caption{Epsilon Data}
\label{fig:xgboost-eps}
\end{subfigure}%
\begin{subfigure}{0.23\textwidth}
\centering
\includegraphics[width=0.98\textwidth]{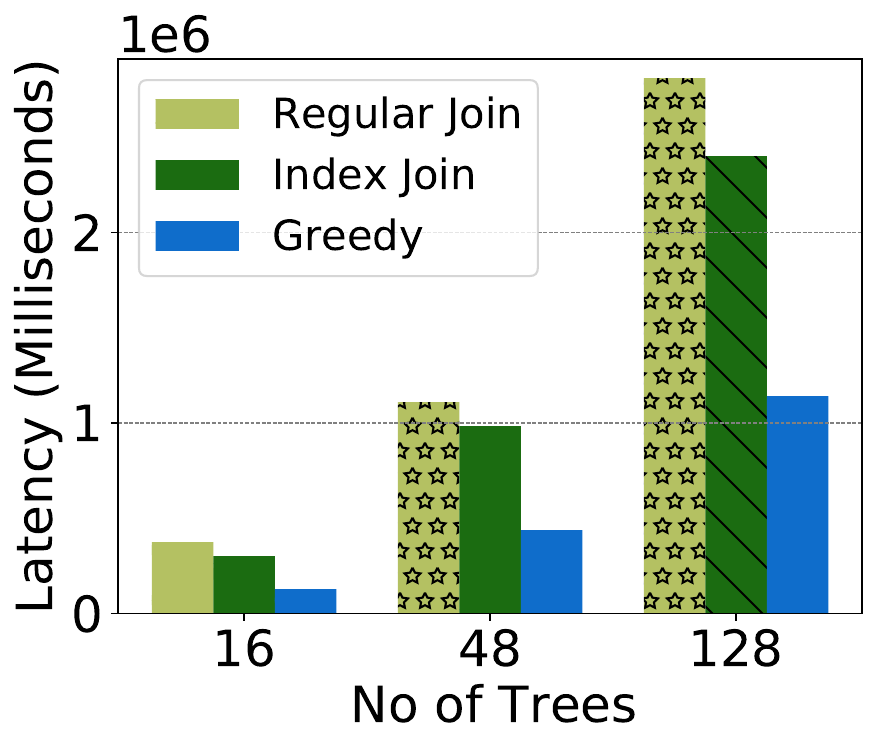}
\caption{Bosch Data}
\label{fig:xgboost-bosch}
\end{subfigure}
\vspace{-8pt}
\caption{Evaluating the Quick Scorer Model}
\label{fig:eval-xgboost}
\vspace{-4pt}
\end{figure}

\begin{figure}[ht]
\begin{subfigure}{0.23\textwidth}
\centering
\includegraphics[width=0.98\textwidth]{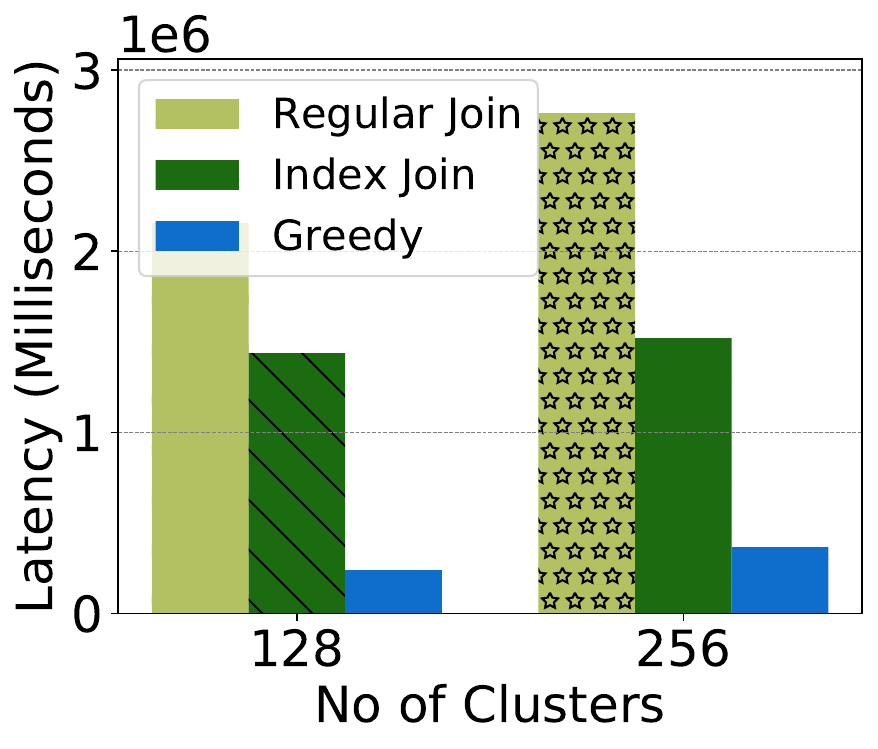}
\caption{Epsilon Data}
\label{fig:pq-eps}
\end{subfigure}%
\begin{subfigure}{0.23\textwidth}
\centering
\includegraphics[width=0.98\textwidth]{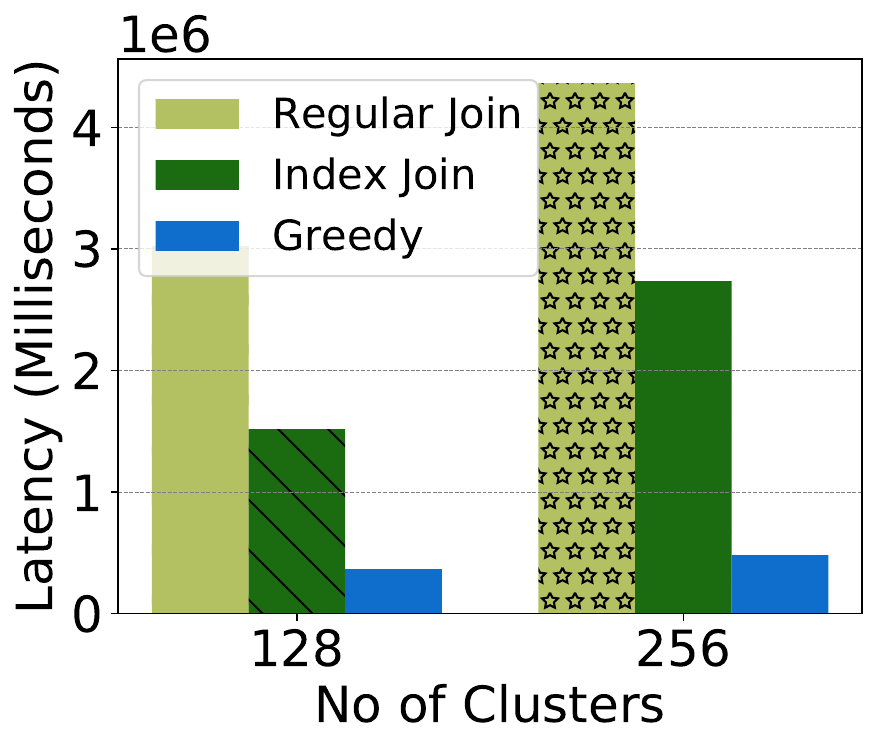}
\caption{Bosch Data}
\label{fig:pq-bosch}
\end{subfigure}
\vspace{-8pt}
\caption{Evaluating Product Quantization}
\label{fig:eval-pq}
\end{figure}

\vspace{2pt}
\noindent
\textbf{Evaluating Quick Scorer and Product Quantization}:
We evaluated the performance of baselines on queries comprising Quick Scorer-based random forest model with Epsilon and Bosch datasets by varying the number of trees in the models. Experimental results indicate that our proposed \textit{Greedy} approach outperforms the regular join and index join by up to $3\times$ and $2.4 \times$ for Epsilon and Bosch, respectively, as illustrated in Figure \ref{fig:eval-xgboost}.
The variations in the number of trees do not significantly impact the speedup because the factorizable model component is computed along the features and does not depend on the number of trees, while the aggregation of outputs from all trees is a non-decomposable operation.

We evaluated the nearest-neighbor search queries comprising product quantization model on Epsilon and Bosch datasets by varying the number of clusters while partitioning the feature vectors into 10 parts. Based on the experimental results, as outlined in Figure \ref{fig:eval-pq}, \textit{Greedy} factorization achieves up to $9\times$ and $6\times$ speedups compared to the regular and index join strategies, respectively. As the number of clusters increases, the number of target vectors for distance computation also increases for each joined tuple. Since \textit{Greedy} factorization computes these distances before performing join, an increase in join cardinality does not affect this step.

Observing the performance of the baselines on multilayer neural networks, quick scorer, and product quantization models, we can state that our \textit{greedy} factorization strategy can be effectively applied in the Python environment also.

\section{Evaluation of Factorization of LLM Inference Queries}
\label{sec:appendix-llm}
We evaluated the benefits of factorization on an LLM workload. This section provides the details of our LLM workload evaluation.

\vspace{3pt}
\noindent
\textbf{Additional Dataset for LLM Query Evaluation}:
Alongside the IMDB dataset, LLM inference evaluation uses the\textit{MovieLens-1M dataset}~\cite{harper2015movielens, sarwat2017database}, which consists of $1$ million ratings provided by $6{,}000$ users for $4{,}000$ movies and $1$ million user ratings.

\subsection{LLM Inference Workload}
We developed three LLM inference queries on IMDB and MovieLens datasets. 
(Q1) \textcolor{black}{joins \textit{user} table with \textit{movie} table on the MovieLens dataset to} pair each user with every movie. Then, in the user-defined inference application, the user and movie information are used to compose LLM prompts for summarizing user profiles and movie details, respectively. These two LLM calls can be factorized from the inference application and pushed down to the user table and movie table, respectively. Then, another LLM call evaluates whether the movie aligns with the user's preferences and explains the recommendation. (Q2) \textcolor{black}{joins \textit{movie info}, \textit{title}, \textit{info type}, and \textit{kind type} tables from the IMDB dataset and} analyzes movie plots through three sequential LLM prompts: (i) Rating each plot's hook quality (1-5 scale), \textcolor{black}{which can be pushed down to the output of the node (J1) that joins movie info and info type tables(, and its ancestor join nodes)}, (ii) Classifying the movie's era appeal (Classic/Modern/Niche), \textcolor{black}{which can be pushed down to the output of the node that joins J1 with title table}, and (iii) Evaluating whether the movie could be reclassified into alternative formats based on title, era, and hook strength.
The third query (Q3) \textcolor{black}{performs three sequential joins on the IMDB dataset across \textit{person info}, \textit{movie info}, and \textit{movie companies} tables}, producing <person, movie, company> triples. These triples are sent to an inference application that invokes LLMs with multiple prompts: (i) Summarizing person, movie, and company profiles separately, which can be factorized from the inference application and pushed down to these three tables; (ii) Determining whether a given person and movie are meaningfully linked, which can be pushed to the output of the query node that joins \textcolor{black}{person info and movie info tables and its ancestor join nodes}; 
and (iii) Inferring whether the person worked with the company via that movie based on the inferred relationship and company history. 
List.~\ref{lst:llm-query} describes the three LLM queries used in our experimental evaluation of LLM inference workload. We utilized OpenAI's RESTful API to invoke the \texttt{gpt-3.5-turbo} model for the LLM calls in all the queries.

\subsection{LLM Inference Evaluation Results}

\begin{figure}[ht]
\centering
\includegraphics[width=0.48\textwidth]{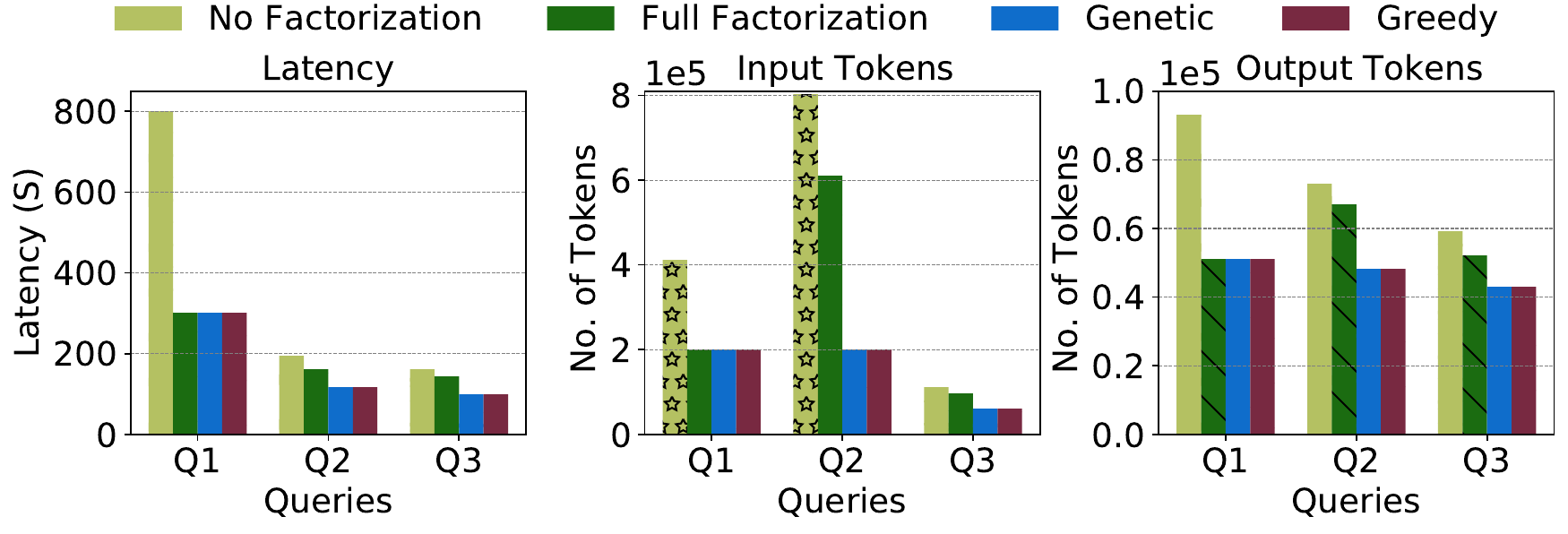}
\vspace{-14pt}
\caption{\small End-to-End Latency Comparison for LLM Queries. (Input/Output tokens refer to the number of tokens received and sent back by the LLM.) }
\label{fig:eval-llm}
\end{figure}

As shown in Fig. \ref{fig:eval-llm}, factorization optimization can reduce the number of tokens processed by the LLM from up to $71\%$, with an end-to-end latency improvement of up to $2.6\times$. 
In LLM inference, the prompt response time dominates over the query joining time, which makes the factorization and push down beneficial to queries of any complexity.
\textcolor{black}{In execution latency of Q1, Q2, and Q3, \textit{Greedy} and \textit{Genetic} optimizers outperform other baselines by up to $2.6\times$, $1.68\times$, and $1.63\times$, respectively. The input token reduction by these optimizers is up to $2.0\times$, $4.0\times$, and $1.83\times$ respectively for three queries, while the output token reduction is up to $1.8\times$, $1.5\times$, and $1.37\times$, respectively.}
In Q1, \textit{Full factorization} is the optimal plan, because all factorized computations should be pushed to table scan nodes, which have the lowest cardinalities compared to intermediate join nodes. For this case, our \textit{Greedy} and \textit{Genetic} optimizers also selected the full factorization plan. \textcolor{black}{In both Q2 and Q3, \textit{Greedy} and \textit{Genetic} optimizers achieved the best performance by selecting the plans that push LLM calls to intermediate join nodes with fewer tuples compared to table scan nodes.}. 

\begin{lstlisting}[style=sqlstyle,caption={LLM Queries}, label={lst:llm-query}] 
// Q1: Using LLM for Movie Recommendation 
SELECT LLM('Please give a recommendation score and explain why', LLM('Please summarize', u.description), LLM('Please summarize', m.description))
FROM user u CROSS JOIN movie m
WHERE m.spoken_language LIKE % English % AND trending_movie_classifier(m.popularity, m.vote_average, m.vote_num) = True;

// Q2: LLM for Classifying Movie Plots
WITH 
movie_base AS (
  SELECT m.movie_id, t.title, t.production_year, SUBSTR(m.plot, 1, 150) AS plot_excerpt, k.kind, CONCAT(CAST(m.movie_id AS VARCHAR), ':', SUBSTR(m.plot, 1, 150)) AS hook_rating_input, CONCAT(t.title, ' (', t.production_year, ')') AS era_type_input FROM ( SELECT mi.movie_id, mi.info AS plot FROM movie_info mi JOIN info_type it ON mi.info_type_id = it.id WHERE it.info = 'plot'
  ) m JOIN titles t ON m.movie_id = t.id CROSS JOIN kind_type k
)
SELECT movie_id, plot_excerpt AS plot, kind, LLM(CONCAT('Movie: ', era_type_input, '; ', 'Hook: ', LLM(hook_rating_input, 'Rate plot hook (1-5) with an explanation of one sentence') AS hook_rating, '; ', 'Era: ', LLM(era_type_input, 'Era appeal: 1- Classic 2- Modern 3- Niche, with explanation') AS era_type, '; ', 'Kind: ', kind), 'Could this work as a [kind]? Consider: 1. Title, 2. Production era, 3. Hook, 4. Era. Answer: Yes/No with explanation') AS reclassification_advice FROM movie_base;

// Q3: LLM for Inferring the Relationship among Movie, Person, and Movie Company
WITH movie_info AS (
  SELECT movie_id, array_agg(info) AS movie_data FROM movies GROUP BY movie_id
),
person_info AS (
  SELECT person_id, array_agg(info) AS person_data FROM persons GROUP BY person_id
),
company_data AS (
  SELECT company_id, array_agg(movie_id) AS company_movies
  FROM movie_companies GROUP BY company_id
),
combined_data AS (
  SELECT p.person_id, m.movie_id, c.company_id, p.person_info, m.movie_info, c.company_movies FROM person_data p CROSS JOIN movie_data m CROSS JOIN company_data c
)
SELECT person_id, movie_id, company_id, LLM('What are the movie companies linked to the person if person and movie are related? Please explain', LLM('Please return whether the person and movie pairs are related to each other and explain why', LLM('Please summarize the person information', person_data), LLM('Please summarize the movie information', movie_data)), LLM('Please summarize the company information', company_movies)) AS person_company_relation
FROM combined_data;

\end{lstlisting}

\end{document}